\documentclass[prd,preprint,preprintnumbers,floatfix,secnumarabic,nofootinbib,amsmath,amsfonts,amssymb]{revtex4}
\usepackage{dcolumn}
\usepackage{bm}
\usepackage{slashed}
\usepackage{graphicx}
\usepackage{epsfig}
\usepackage{pxfonts}
\usepackage[colorlinks=true,citecolor=blue,linkcolor=red,hyperfootnotes=false]{hyperref}
\DeclareMathAlphabet \mathscr{T1}{hlcw}{m}{it}
\newcommand{\SU}{\mathrm{SU}}

\begin{document}
\title{The relic density of shadow dark matter candidates}
\author{Mehrdad~Adibzadeh}
\email{mehrdad@mailaps.org}
\affiliation{Department of Physics, University of Virginia, P.O.Box 400714,
Charlottesville, VA 22904, USA}
\author{P.~Q.~Hung}
\email{pqh@virginia.edu}
\affiliation{Department of Physics, University of Virginia, P.O.Box 400714,
Charlottesville, VA 22904, USA}


\begin{abstract}
We present the results of relic density calculations for cold dark matter
candidates
coming from a model of dark energy and dark matter, which is described by
an asymptotically free gauge group $\SU(2)_Z$ (QZD) with a
coupling constant $\alpha_Z \sim$ 1 at very low scale of $\Lambda_Z \sim
10^{-3}$ eV while $\alpha_Z \sim$ weak coupling at high energies. The dark
matter candidates of QZD are two fermions in the form of weakly interacting
massive particles. Our results show that for masses between 50 and 285 GeV, they
can account for either a considerable fraction or the entire dark matter of the
Universe.
\end{abstract}

\maketitle
\section{Introduction} \label{sec:intro}
It is almost universally accepted that the picture of the Universe made up of 
approximately $4\%$ baryonic matter, $23\%$ dark matter and $73\%$ dark energy
represents a realistic cosmological model. However, it is astounding that almost $96\%$
of the energy density of the Universe resides in some as-yet-unknown form.
What is ``dark matter''? What is ``dark energy''?

In Refs.~\cite{Hung2005,Hung2006a}, a model of dark energy and dark matter was
proposed in which a new unbroken gauge group $\SU(2)_Z$ -- the shadow sector -- grows strong at a scale $\sim 10^{-3}\,$eV. The gauge group $\SU(2)_Z$ was nicknamed Quantum Zophodynamics, or QZD, in  Refs.~\cite{Hung2005,Hung2006a}, where the subscript ``\textit{Z}" stands for the Greek word \textit{Zophos}, meaning
darkness. The model is described by an $\SU(2)_Z$ instanton-induced
potential of an axion-like particle, $a_Z$, which possesses two degenerate
minima. The degeneracy is lifted by a mechanism described in
Refs.~\cite{Hung2006a,Hung2007a}, yielding a false vacuum with energy density $\sim (10^{-3}\,\text{eV})^4$ and a true
vacuum with vanishing energy density. The present Universe is assumed to be trapped in the false
vacuum~\cite{Hung2007b}, whose energy density mimics the cosmological constant. This is, in a nutshell, the
dark energy model proposed in Ref.~\cite{Hung2006a}, which also computed various quantities
of interest such as the tunneling rate to the true vacuum, etc. 
A Grand Unified Theory (GUT) involving the SM and $\SU(2)_Z$ was considered by
Ref.~\cite{Hung2006d} (The models presented in Refs.~\cite{Hung2006a,Hung2006d} were later 
revisited by Refs.~\cite{Das}.). 

The particle content of the model includes two shadow fermions, $\psi _{\left(
{L,R} \right),i}^{\left( Z \right)} $ with $i=1,2$,
which transform as $\left( {1,1,0,3} \right)$ under $\SU(3)_c \otimes \SU(2)_L
\otimes \mathrm{U}(1)_Y \otimes \SU(2)_Z$,
two messenger scalar fields (mediating between the QZD and SM matters; one of which is much heavier than the other
\cite{Hung2006a})  $\tilde{\bm{\varphi}} _i^{\left( Z \right)} $ with $i=1,2$ 
transforming as $\left( {1,2,Y_{\tilde \varphi}   =  - 1,3} \right)$, and one
singlet complex scalar field $\phi _Z  = \left( {1,1,0,1} \right)$ whose imaginary part
plays the role of the axion-like particle mentioned above.

As discussed in Ref.~\cite{Hung2006a}, the masses of the $\SU(2)_Z$ triplet
shadow fermions are found to be of the order of 100 - 200 GeV for the $\SU(2)_Z$ gauge coupling to grow strong
at a scale $\sim 10^{-3}$ eV, needed for the dark energy scenario. This coupling
constant starts out at GUT-scale energy with a value comparable to that of the electroweak couplings,
remains relatively flat until an energy comparable to the shadow fermion masses is
reached, and then  starts to grow after the shadow fermions drop out of the Renormalization Group
(RG) equations. At that dropout point, the $\SU(2)_Z$ gauge coupling becomes comparable to the
weak $\SU(2)_L$ coupling at the electroweak scale energy.
These features have interesting consequences concerning the possibility of
the shadow fermions being candidates for cold dark matter (CDM) in
the form of weakly interacting massive particles (WIMP's)
\footnote{For a review on various features of CDM and WIMP, see, e.g., Refs.
\cite{CDMRevs}.}. The main reason is the fact that
the annihilation cross sections for two shadow fermions into two $\SU(2)_Z$
``shadow gluons'' are of the order of the weak cross sections, a typical requirement for
WIMP's. The estimates that were made in
Ref.~\cite{Hung2006a} showed that it was possible for shadow fermions to be candidates for
CDM with the \textit{right} relic density.

In this work, we would like to investigate this scenario in more details and by solving shadow fermions'
evolution equations to determine the conditions under which they can be considered to be WIMP 
cold dark matter candidates. It will be seen that the mass range for the shadow
fermions obtained by the requirement of having the \textit{right} density fits
in snugly with that used in the RG equations (i.e., the $\SU(2)_Z$ gauge coupling grows strong
at a scale $\sim 10^{-3}$ eV).

The outline of the paper is as follows. First, we go over the QZD model as far
as the issue of dark matter is concerned. Then, we derive the evolution
equations for shadow fermions and consequently solve them numerically, to obtain
their relic density. Finally, the results of our relic density calculations will
be presented and discussed, in comparison with the observational values.
The shadow fermions relic density, when computed, would only depend on their
masses. Therefore, the parameter space is simply two dimensional.
\section{The shadow sector and its candidates for cold dark matter}
In this work, we only concentrate on the potential candidates for cold dark matter that QZD provides in the form of
fermions. However, as discussed in Refs.~\cite{Hung2005,Hung2006a}, the model
offers a mechanism for leptogenesis through the decay of a messenger field,
resulting in a net SM lepton surplus.

For clarity, we list the particle content that is useful
for our calculations, in particular the transformation of these
particles under $\SU(3)_c \otimes \SU(2)_L \otimes \mathrm{U}(1)_Y \otimes \SU(2)_Z$.
\begin{itemize}
\item Two shadow fermions $\psi _{\left(
{L,R} \right),i}^{\left( Z \right)} $ with $i=1,2$, which transform as $\left(
{1,1,0,3} \right)$.
\item  Two messenger scalar fields $\tilde{\bm{\varphi}} _i^{\left( Z
\right)} $, with $i=1,2$, transforming as $\left( {1,2,Y_{\tilde \varphi}   =  -
1,3} \right)$. For relic density calculations, only the one with
mass $\mathcal{O}(<1\,\text{TeV})$, i.e.,  $\tilde{\bm{ \varphi}} _1^{\left( Z
\right)}$, plays a role while the very heavy one with
GUT-scale mass, i.e.,  $\tilde{\bm{ \varphi}} _2^{\left( Z \right)}$, is only
useful for leptogenesis in this picture~\cite{Hung2006b}.
\item One singlet
complex scalar field $\phi _Z  = \left( {1,1,0,1} \right)$.
The imaginary part $a_Z$ plays the role of the axion-like particle
mentioned in section~\ref{sec:intro}. The real part, $\sigma_Z$, was used
as the inflaton in a model of ``low-scale'' inflationary
universe~\cite{Hung2006c}. 
\end{itemize}

We now briefly review the relevant aspects of the shadow sector that
would be used in our relic density calculations for shadow fermions.
\subsection{The QZD Lagrangian}
The Lagrangian of $G_{\text{SM}} \otimes \SU(2)_Z$ is given by~\cite{Hung2006a}
\begin{equation}\label{equ:Lag}
\mathcal{L} = \mathcal{L}_{\text{SM}}  + \mathcal{L}_{\text{kin}}^Z  +
\mathcal{L}_{\text{Yuk}}  + \mathcal{L}_{CP}  - V\left( {\left| {\tilde{
\bm{\varphi}} ^{\left( Z \right)} } \right|^2 } \right) - V\left( {\left| {\phi
_Z } \right|^2
} \right) \, , 
\end{equation}
where $\mathcal{L}_{\text{SM}}$ is the SM Lagrangian and
\begin{subequations}\label{equ:Lags}
\begin{equation}\label{equ:Lagkin}
\mathcal{L}_{\text{kin}}^Z  =  - \frac{1}
{4}\bm{G}_{\mu \nu }^{\left( Z \right)}  \cdot \bm{G}^{\left( Z \right),\mu \nu
}  + \frac{1}
{2}\sum\limits_i {\left| {D_\mu  \tilde{\bm{ \varphi}} _i^{\left( Z \right)} }
\right|^2 } +
i\sum\limits_j {\bar \psi _{\left( {L,R} \right),j}^{\left( Z \right)}
\slashed{D} \psi _{\left( {L,R} \right),j}^{\left( Z \right)} } \, ,
\end{equation}
\begin{equation}\label{equ:LagYuk}
\mathcal{L}_{\text{Yuk}}  = \sum\limits_i {\sum\limits_m {\left( {g_{\tilde
\varphi _1 m}^i \bar l_L^m \tilde{\bm{\varphi}} _1^{\left( Z \right)} \psi
_{i,R}^{\left( Z \right)}  + g_{\tilde \varphi _2 m}^i \bar l_L^m
\tilde{\bm{\varphi}} _2^{\left( Z \right)} \psi _{i,R}^{\left( Z \right)} }
\right)} }  + \sum\limits_i {K_i \bar \psi _{i,L}^{\left( Z \right)} \phi _Z
\psi _{i,R}^{\left( Z \right)} }  +\text{ H.c.}\, ,
\end{equation}
\begin{equation}\label{equ:LagCP}
\mathcal{L}_{CP}  = \frac{{\theta _Z }}
{{32\pi ^2 }}\bm{G}_{\mu \nu }^{\left( Z \right)}  \cdot \tilde{\bm{G}}^{\left(
Z \right),\mu \nu } .
\end{equation}
\end{subequations}

In the above Lagrangians, $G_{\mu \nu}^{\left( Z \right)}$'s are the
field-strength tensors of
$\SU(2)_Z$ gauge bosons, the so-called shadow gluons, and the boldface typeset
indicates the $\SU(2)_Z$ triplet multiplicity. The sum over $m$ is in fact over
the number of SM families and the summation over $i$ includes the number of
shadow fermions. The coefficients $g_{\tilde \varphi _1 m}$, $g_{\tilde \varphi
_2 m}$, and $K_i $ are complex. The covariant derivative in the Lagrangian can be written in the
form
\[
D_\mu   = \partial _\mu   - i\frac{g}
{2}\hat{\bm{\tau}}  \cdot \bm{W}_\mu   - i\frac{{g'}}
{2}\hat YB_\mu   - ig_Z \hat{\bm{T}} \cdot \bm{A}_\mu ^{\left( Z \right)} ,
\]
where $\hat{T}$'s are the generators of $\SU(2)_Z $, which ought to be in
adjoint representation when acting on shadow fermions, and $A_\mu^{\left( Z
\right)}$'s are the shadow gluon fields.
The QZD Lagrangian is invariant under a $\mathrm{U}(1)_A^{\left({Z}\right)}$
global symmetry, which yields an instanton-induced axion-like potential driving
the present accelerating Universe. The transformations of QZD and SM particles
under this $\mathrm{U}(1)_A^{\left({Z}\right)}$ global symmetry is given in
detail in Ref.~\cite{Hung2006a}.
\subsection{Masses and coupling constant}
The masses of shadow fermions come from the spontaneous breakdown of
$\mathrm{U}(1)_A^{\left({Z}\right)}$. Such a breakdown is made possible through
the vacuum expectation value of $\phi_Z$. Therefore, in the Yukawa coupling of
shadow fermions with $\phi_Z$, given in Eq.~(\ref{equ:LagYuk}), when $\phi_Z$
attains vacuum expectation value, $\left\langle \phi_Z \right\rangle  = v_Z$,
shadow fermions receive masses
\begin{subequations}
 \begin{equation}
 m_{\psi_1^{\left({Z}\right)}} = \left|{K_1}\right| v_Z \, ,
\end{equation}
 \begin{equation}
 m_{\psi_2^{\left({Z}\right)}} = \left|{K_2}\right| v_Z \, .
\end{equation}
\end{subequations}

The scalar messenger fields, on the other hand, are assumed to have zero vacuum
expectation values to keep QZD symmetry unbroken. 
Their masses are non-trivially constrained by the evolution of QZD coupling,
as explained in Ref.~\cite{Hung2006a}.

The QZD coupling constant, $\alpha _Z  = {{g_Z^2 } \mathord{\left/ {\vphantom
{{g_Z^2 } {4\pi }}} \right. \kern-\nulldelimiterspace} {4\pi }}$, is close to
the SM couplings at high energies, while it increases to $\alpha _Z  \sim 1$ at
$\Lambda_Z \sim 3 \times 10 ^{-3}$ eV. The RG analysis of $\alpha_Z$, conducted
in Ref.~\cite{Hung2006a}, studies the evolution of $\alpha_Z$ from
$M_{\text{GUT}}$ to $\Lambda_Z $ through a two-loop $\beta$ function for
possible masses of QZD particles. 

The RG analysis results indicate a direct correlation between the scale at which
$\alpha_Z \left({E}\right)$ starts increasing promptly and the mass of the
lighter shadow fermion, $m_1$. At energies prior to $m_1$, $\alpha_Z
\left({E}\right)$ is mostly flat, but upon $E \sim m_1$ it begins to grow toward
its value at $\Lambda_Z$, i.e., $\alpha _Z \left( {\Lambda _Z } \right) \sim 1$.

Ref.~\cite{Hung2006a} provides $\alpha_Z \left({E}\right)$ values for different
conditions, i.e., masses, number of messenger fields, etc. However, a common
thread among all analyses is that $\alpha_Z$ does not change much from its value
at $M_{\text{GUT}}$ until $E \sim m_1$, being almost scale independent in that
interval. At energies comparable to the masses of the shadow fermions,
which themselves are of the order of he electroweak scale, $\alpha_Z$
is comparable to the electroweak $\SU(2)_L$ gauge coupling.
This will partially qualify QZD's shadow fermions as WIMP's and their
candidacy for CDM, as already explained.
\subsection{Shadow fermions as candidates for cold dark matter}
The two shadow fermions of QZD particle content meet the criteria for a WIMP,
since
\begin{itemize}
\item
They interact very weakly with normal matter, i.e., through heavy scalar fields~\cite{Hung2006a}.
\item
They have cross sections of weak strength: masses in GeV and coupling
constant in order of weak coupling~\cite{Hung2006a}.
\item
At least one is stable on cosmological scales: The lighter of the two shadow
fermions is stable. The heavier one can decay into SM leptons and the lighter
shadow fermion through the messenger scalar field
(see Appendix~\ref{app:dec}). However, if the shadow fermion masses are
degenerate, both can
be stable. Additionally, the shadow fermions can annihilate into shadow gluons
or each other (if kinematically allowed).
\end{itemize}

The messenger fields do not qualify as CDM candidates
since they are unstable. The relic densities of shadow fermions can be obtained
reliably by solving their evolution equations. Solving the evolution equations
will reveal the applicable masses, which would give meaningful relic densities
and put the model's candidates for dark matter into the test.
\section{Evolution equations for shadow fermions}
The standard Boltzmann equation~\cite{Kolb1990} describing the evolution of the
number density $n$ of a particle species $\psi$, is
\begin{equation}\label{equ:Sbolt}
\frac{{dn}}
{{dt}} + 3Hn =  - \left\langle {\sigma v} \right\rangle \left( {n^2  -
n_{{\text{eq}}}^2 } \right),
\end{equation}
where $H$ is the Hubble parameter, $n_{\mathrm{eq}}$ is the equilibrium density,
$v$ is the relative velocity in the annihilation process $\psi \bar \psi  \to
{\text{all}}$, and $\left\langle {\sigma v} \right\rangle $ denotes the thermal averaging of
$\sigma v$, with $\sigma $ being the total cross section of the annihilation
reaction. The equilibrium density $n_{\text{eq}}$ is given by
\begin{equation}\label{equ:Genneq}
 n_{{\text{eq}}}  = \frac{g}
{{\left( {2\pi } \right)^3 }}\int {d^3 {\mathbf{p}}} f\left(
{{\mathbf{x}},{\mathbf{p}}} \right),
\end{equation}
where $g$ is the species internal degrees of freedom and $f\left(
{{\mathbf{x}},{\mathbf{p}}} \right)$ is the equilibrium distribution function.
For particles that may play the role of CDM, the equilibrium number density in
the nonrelativistic approximation is
\[
n_{{\text{eq}}}  \approx g\left( {\frac{{mT}}
{{2\pi }}} \right)^{\tfrac{3}
{2}} e^{ - \frac{m}
{T}} ,
\]
where $T$ is the temperature, and $m$ is the mass of the relic. The number
density $n$ satisfying Eq.~(\ref{equ:Sbolt}) has two behaviors. In early times,
$n$ closely follows $n_{\mathrm{eq}}$ but later when the temperature drops below
$m$, the mass of the species, $n_{\mathrm{eq}}$ starts to decrease exponentially
until a``freeze-out" temperature is reached where the annihilation rate is not
fast enough to maintain equilibrium. Below this temperature, $n$ deviates
substantially from $n_{\mathrm{eq}}$ and eventually gives the present day
abundance of the species.
Equation~(\ref{equ:Sbolt}) can be solved numerically in relativistic (hot relic)
 or nonrelativistic (cold relic) regime.
Ref.~\cite{coann} showed that the validity of
Eq.~(\ref{equ:Sbolt}) and its solution breaks down if the relic particle is the
lightest of a set of particles whose masses are near-degenerate and can
contribute to the density of the relic through annihilation or decay processes,
the so-called coannihilation case.

For QZD's cold dark matter candidates, both shadow fermions can have present day
abundances, if they have similar masses, which blocks the decay channel. For
that
reason, the evolution equations for the number densities of both species ought
to be
considered. The trivial reduction of shadow fermions 
occurs through their annihilations into QZD gauge bosons and the decay of the
heavier one. Parallel to that, shadow fermions can annihilate into each other as
well, which is analogous to the coannihilation case of Ref.~\cite{coann}.

To summarize, the reactions entering into Boltzmann equations for densities of
shadow fermions are
\begin{itemize}
 \item 
Annihilation of shadow fermions into shadow gluons:  $\psi _{i}^{\left( Z
\right)} \bar \psi _{i}^{\left( Z \right)}  \rightleftarrows
\bm{A}^{\left({Z}\right)}\bm{A}^{\left({Z}\right)}$
 \item 
Annihilation of a pair of one species into a pair of another: $\psi _1^{\left( Z
\right)} \bar \psi _1^{\left( Z \right)}  \rightleftarrows \psi _2^{\left( Z
\right)} \bar \psi _2^{\left( Z \right)} $. 
 \item 
The decay of the heavier one into the lighter one and SM leptons: $\psi
_2^{\left( Z \right)} \rightarrow l\bar{l'} \psi _1^{\left( Z \right)} $. 
\end{itemize}

We assume negligible chemical potential for shadow fermions, which implies
symmetry among the number densities for particle and antiparticle of each
species. To be inclusive, there can be a particle-antiparticle asymmetry in the shadow sector originating from the decay mechanism of messenger fields. The decay of messenger fields induces a particle-antiparticle asymmetry in SM leptons (see the decay of a messenger boson in Fig.~\ref{fig:decdiag}). The corresponding asymmetry in the shadow sector is expected to be as small as $\mathcal{O} \left({10^{-7}}\right)$  and therefore negligible to be considered in our relic density calculations.

The evolution equations for number densities $n_1$, $n_2$ of shadow fermions
$\psi_1^{(Z)}$, $\psi_2^{(Z)}$ are in the form
\begin{subequations}\label{equ:n12eq}
\begin{equation}\label{equ:n1eq}
\frac{{dn_1 }}
{{dt}} + 3Hn_1 =  - \frac{1}
{2}\left\langle {\sigma_{1A} v_{1A}} \right\rangle \left(
{n_1^2  - n_{1,\mathrm{eq}}^2 } \right) - \frac{1}
{2}\left\langle {\sigma _{12} v_{12} } \right\rangle n_1^2  + \frac{1}
{2}\left\langle {\sigma _{21} v_{21} } \right\rangle n_2^2 \, ,
\end{equation}
\begin{equation}\label{equ:n2eq}
\frac{{dn_2 }}
{{dt}} + 3Hn_2 =  - \frac{1}
{2}\left\langle {\sigma_{2A} v_{2A}} \right\rangle \left(
{n_2^2  - n_{2,\mathrm{eq}}^2 } \right) - \frac{1}
{2}\left\langle {\sigma _{21} v_{21} } \right\rangle n_2^2  + \frac{1}
{2}\left\langle {\sigma _{12} v_{12} } \right\rangle n_1^2 -\Gamma_{21} \left(
{n_2  - n_{2,\mathrm{eq}}} \right)\, ,
\end{equation}
\end{subequations}
where $\Gamma_{21}$ is the decay rate of the heavier shadow fermion, i.e., $\psi
_{2}^{\left( Z \right)}$, $\sigma_{ij}$ (with $i,j=1,2,A$) refers to
the total annihilation cross section for the processes
\begin{subequations}\label{equ:reactions}
\begin{equation}\label{equ:12togg}
\psi _{i}^{\left( Z \right)} \bar \psi _{i}^{\left( Z \right)}  \longrightarrow
\bm{A}^{\left({Z}\right)}\bm{A}^{\left({Z}\right)},
\end{equation}
\begin{equation}\label{equ:1to2}
 \psi _i^{\left( Z \right)} \bar \psi _i^{\left( Z \right)} \longrightarrow \psi
_j^{\left( Z \right)} \bar \psi _j^{\left( Z \right)},
\end{equation}
\end{subequations}
and $v_{ij}$ is the relative velocity of the annihilating particles for each
reaction. Also, with a Maxwell-Boltzmann distribution function\footnote{It has
been shown that the use of correct statistics would only amount to less than 1\%
difference (see Ref.~\cite{Scherrer1986}).}, $n_{i,{\text{eq}}}$ is given by
\begin{align}\label{equ:neq}
 n_{i,{\text{eq}}}  &= \frac{g_i}
{{\left( {2\pi } \right)^3 }}\int {d^3 {\mathbf{p}}} e^{-{E_i \mathord{\left/
{\vphantom {E_i T_Z}} \right. \kern-\nulldelimiterspace} T_Z}} \cr
&= \frac{T_Z}{2\pi^2 } g_i m_i^2 K_2\left({\frac{m_i}{T_Z }}\right)\, ,
\end{align}
where $T_Z$ is the temperature of QZD matter, $m_i$ is the mass of the species
and $K_2$ is the modified Bessel function of second kind. The ${1
\mathord{\left/ {\vphantom {1 2}} \right. \kern-\nulldelimiterspace} 2}$ factor
on the right hand side of Eqs.~(\ref{equ:n12eq}) is to account for non-identical
annihilating shadow fermions.

Equations ~(\ref{equ:n12eq}) can be written in a more convenient form by
considering the number of particles in a comoving volume
\begin{equation}
Y_i  = \frac{{n_i }}
{s} \, ,
\end{equation}
which is the ratio of number density to entropy density, with the time
derivative in the form
\begin{equation}
\frac{{dY_i }}
{{dt}} = \frac{1}
{s}\frac{{dn_i }}
{{dt}} - \frac{{n_i }}
{{s^2 }}\frac{{ds}}
{{dt}} \, .
\end{equation}
In the absence of entropy production (i.e., $s = {S \mathord{\left/ {\vphantom
{S {R^3 }}} \right. \kern-\nulldelimiterspace} {R^3 }}$ with $S =
\mathrm{const.}$)
\begin{equation}
\frac{{ds}}
{{dt}} =  - 3\frac{{S}}
{{R^3 }}\frac{1}
{R}\frac{{dR}}
{{dt}} =  - 3Hs \, ,
\end{equation}
which results in
\begin{equation}
s\frac{{dY_i }}
{{dt}} = \frac{{dn_i }}
{{dt}} + 3Hn_i \; .
\end{equation}
The evolution equations, then, can be reformulated in the form
\begin{subequations}\label{equ:Y12eq}
\begin{equation}\label{equ:Y1eq}
\frac{{dY_1 }}
{{dt}}=  \frac{s}
{2} \left[-{\left\langle {\sigma_{1A} v_{1A}} \right\rangle
\left( {Y_1^2  - Y_{1,\mathrm{eq}}^2 } \right) - \left\langle {\sigma _{12}
v_{12} } \right\rangle Y_1^2  + \left\langle {\sigma _{21} v_{21} }
\right\rangle Y_2^2}\right] ,
\end{equation}
\begin{equation}\label{equ:Y2eq}
\frac{{dY_2 }}
{{dt}}=   \frac{s}
{2}\left[{-\left\langle {\sigma_{2A} v_{2A}} \right\rangle
\left( {Y_2^2  - Y_{2,\mathrm{eq}}^2 } \right) - \left\langle {\sigma _{21}
v_{21} } \right\rangle Y_2^2  + \left\langle {\sigma _{12} v_{12} }
\right\rangle Y_1^2 -\frac{2}{s}\Gamma_{21} \left( {Y_2  - Y_{2,\mathrm{eq}}}
\right)}\right],
\end{equation}
\end{subequations}
where $Y_{i,\text{eq}}=n_{i,\text{eq}}/s$. Additionally, it is convenient to use
the QZD plasma temperature $T_Z$ as independent variable, in place of time $t$. 
The relation between $T$ (the photon temperature) and $T_Z$ is easily found by
the entropy conservation \cite{Hung2005,Hung2006a}. The technique is essentially
the same as that for
finding the neutrino temperature using entropy conservation~\cite{Kolb1990}. For
example, at temperatures higher than the mass of the lighter messenger field
(i.e., $\tilde{\bm{ \varphi}} _1^{\left( Z \right)}$)
$T > m_{\varphi_1} $, the QZD matter is in thermal equilibrium with the rest of
the Universe, i.e., $T_Z=T$. When $T$ falls below the mass of the lighter 
messenger field, $T < m_{\varphi_1} $, the QZD plasma conserves its own entropy
separately and maintains its own temperature $T_Z \ne T$. The relation between
$T$ and $T_Z$ from there on can be found by entropy conservation anytime a
particle decouples and transfers its entropy to the relativistic matter. At
present, i.e., after $e^\pm$ decoupling, $T_Z  = \left[ {{{\left( {{{43}
\mathord{\left/ {\vphantom {{43} {583}}} \right. \kern-\nulldelimiterspace}
{583}}} \right)} \mathord{\left/ {\vphantom {{\left( {{{43} \mathord{\left/
{\vphantom {{43} {583}}} \right. \kern-\nulldelimiterspace} {583}}} \right)}
{\left( {{4 \mathord{\left/ {\vphantom {11 {18}}} \right.
\kern-\nulldelimiterspace} {18}}} \right)}}} \right. \kern-\nulldelimiterspace}
{\left( {{11 \mathord{\left/ {\vphantom {11 {18}}} \right.
\kern-\nulldelimiterspace} {18}}} \right)}}} \right]^{{1 \mathord{\left/
{\vphantom {11 3}} \right. \kern-\nulldelimiterspace} 3}} T$.
Ref.~\cite{Hung2005,Hung2006a} discusses the relation between $T_Z$ and $T$ in
more
detail. Let us define $x_i  = {{m_i } \mathord{\left/ {\vphantom {{m_i } {T_Z
}}} \right. \kern-\nulldelimiterspace} {T_Z }}$, we have
\begin{equation}
\frac{{dY_i }}
{{dt}} = \frac{{dY_i }}
{{dx_i }}\frac{{dx_i }}
{{dt}} =  - \frac{{dY_i }}
{{dx_i }}\frac{{m_i }}
{{T_Z^2 }}\frac{{dT_Z }}
{{dt}}\, ,
\end{equation}
where the time derivative of $T_Z$ satisfies
\begin{equation}
\left( {\frac{{dT_Z }}
{{dt}}} \right)^{ - 1}  = \frac{1}
{{3Hs}}\frac{{x_i^2 }}
{{m_i}}\frac{{ds}}
{{dx_i }}\, .
\end{equation}
Considering all this, we can rewrite Eqs.~(\ref{equ:Y12eq}) in their final forms
\begin{subequations}\label{equ:Y12eqx}
\begin{equation}\label{equ:Y1eqx}
\frac{{dY_1 }}
{{dx_1}}=  \frac{x_1}
{6H} \frac{{ds}}
{{dx_1 }} \left[{ \left\langle {\sigma_{1A} v_{1A}}
\right\rangle \left( {Y_1^2  - Y_{1,\mathrm{eq}}^2 } \right) + \left\langle
{\sigma _{12} v_{12} } \right\rangle Y_1^2  - \left\langle {\sigma _{21} v_{21}
} \right\rangle Y_2^2}\right] ,
\end{equation}
\begin{equation}\label{equ:Y2eqx}
\frac{{dY_2 }}
{{dx_2}}=  \frac{x_2}
{6H} \frac{{ds}}
{{dx_2 }} \left[{ \left\langle {\sigma_{2A} v_{2A}}
\right\rangle \left( {Y_2^2  - Y_{2,\mathrm{eq}}^2 } \right) + \left\langle
{\sigma _{21} v_{21} } \right\rangle Y_2^2  - \left\langle {\sigma _{12} v_{12}
} \right\rangle Y_1^2 +\frac{2}{s}\Gamma_{21} \left( {Y_2  - Y_{2,\mathrm{eq}} }
\right)}\right] .
\end{equation}
\end{subequations}

Equations~(\ref{equ:Y12eqx}) are first-order coupled differential equations in
the form of Riccati equation, which ought to be solved numerically. The
integration of Eqs.~(\ref{equ:Y12eqx}) from early Universe to present
$T_Z^0=1.346$ K (corresponding to photon temperature $T = 2.725$ K) yields
today's number densities $Y_i^0$. The present-day relic density of shadow
fermion $\psi_i^{\left({Z}\right)}$ in units of critical density
$\rho_{\text{crit}}$ is then
\begin{equation}\label{equ:omega}
 \Omega_i= \frac{\rho_{\psi_i^{\left({Z}\right)}}}{\rho_{\text{crit}}}=\frac{s_0
m_i Y_i^0}{\rho_{\text{crit}}}\;,
\end{equation}
where $s_0$ is the present-day entropy density of the shadow sector and
$\rho_{\text{crit}}=3H_0^2/8 \pi G$. Finally, with $H_0 = 100
h$~$\text{km}~\text{sec}^{-1}~\text{Mpc}^{-1}$ and $s_0 = 12
\pi^2T_Z^{0\,3}/45$, Eq.~(\ref{equ:omega}) can be written in the from
\begin{equation}\label{equ:omegah2}
 \Omega_i h^2= 0.5080 \times 10^8 \frac{m_i}{\text{GeV}} Y_i^0\;,
\end{equation}
where $h$ is the Hubble constant in units of
100~$\text{km}~\text{sec}^{-1}~\text{Mpc}^{-1}$. Since
$\psi_2^{\left({Z}\right)}$ decays, the relevant relic density is that of
$\psi_1^{\left({Z}\right)}$. If $m_1= m_2$, however, both shadow fermions can
have present day abundances and only in such case, may we speak of two relic
densities.
\section{Computational method}
Equations~(\ref{equ:Y12eqx}) include thermal averages $\langle{\sigma
v}\rangle$'s, Hubble parameter $H$, and the derivative of entropy density ${{ds}
\mathord{\left/ {\vphantom {{ds} {dx_i }}} \right. \kern-\nulldelimiterspace}
{dx_i }}$, all of which need to be determined for numerical integration.

The annihilation cross sections and the decay rate $\Gamma _{21}$ can be
calculated analytically. They are derived in Appendixes ~\ref{app:ACS} and
\ref{app:dec} and are given in closed forms, to leading order. The thermal
averages $\langle{\sigma v}\rangle$'s were then computed numerically using the
compact integral form of Ref.~\cite{Gondolo1991}. In
Appendix~\ref{app:TA}, the relativistic thermal averages are provided in closed
integral forms, expressed in terms of $x_i$.

On the other hand, the Hubble parameter in a radiation-dominated Universe is
given by 
\begin{equation}\label{equ:Hub}
H = \sqrt {\frac{8}
{3}\pi G\rho } ,
\end{equation}
where $G$ is the gravitational constant and $\rho$ is the total energy density
of the Universe, written as
\begin{equation}\label{equ:rho}
\rho  = g_{\text{eff}} \left( T \right)\frac{{\pi ^2 }}
{{30}}T^4 ,
\end{equation}
where $g_{\text{eff}} \left( T \right)$ is the effective number of relativistic
degrees of freedom.
Ref.~\cite{Gondolo1991} provides $g_{\text{eff}} \left( T \right)$ values for
two QCD phase transition temperatures $T_{\text{QCD}}= 150$ and 400 MeV. We made
use of the $g_{\text{eff}} \left( T \right)$ values corresponding to
$T_{\text{QCD}}=150$ MeV, which is a smoother function, as opposed to
$T_{\text{QCD}}=400$ MeV. It turns out that the solutions to
Eqs.~(\ref{equ:Y12eqx}) do not depend on the choice of $T_{\text{QCD}}$, mainly
because the freeze-out temperatures for shadow fermions are always much higher
than $T_{\text{QCD}}$, due to their large masses. As we already discussed, the
relation between $T$ and $T_Z$ can be easily determined  by entropy
conservation. As a result, the Hubble parameter in evolution equations was
evaluated in terms of $T_Z$ and consequently $x_i$, consistently.

The entropy density $s$ , in Eqs.~(\ref{equ:Y12eqx}), is \textit{mostly} the
entropy of the shadow sector. For temperatures $T>m_{\varphi_1}$, the QZD matter
is in thermal equilibrium with normal matter and $s$ is 
\begin{equation}
 s=\frac{2\pi^2}{45} g_{\ast s} T^3,
\end{equation}
where $g_{\ast s}= {{459} \mathord{\left/ {\vphantom {{459} {4}}} \right.
\kern-\nulldelimiterspace} {4}}$, and  $T=T_Z$. However, for most of the time
$T<m_{\varphi_1}$ and $s$ is the entropy of the shadow sector, which is
conserved
separately, given by
\begin{equation}\label{equ:ent}
s = \frac{{2\pi ^2 }}
{{45}}\left[ {\sum\limits_{{\text{Bosons}}} {g_B T_Z^3 }  + \frac{7}
{8}\sum\limits_{{\text{Fermions}}} {g_F T_Z^3 } } \right].
\end{equation}
In both cases $s$ is easily evaluated in terms of $x_i$, providing values for
${{ds} \mathord{\left/ {\vphantom {{ds} {dx_i }}} \right.
\kern-\nulldelimiterspace} {dx_i }}$ of Eqs.~(\ref{equ:Y12eqx}).

The numerical integration of  the density evolution equations,
Eqs.~(\ref{equ:Y12eqx}), was carried out using an implicit trapezoidal
scheme\footnote{We implemented the idea of the backward differentiation formulas
adapted to implicit trapezoidal scheme, presented in Ref.~\cite{Dieci1992}, for
a
system of Riccati equations.}. We integrate from $x_i=0$ to $x_i= m_i/T_Z^0$,
where $T_Z^0 = 1.346$ K  is the present-day temperature of the QZD matter
corresponding to $T_0 = 2.725$ K, the photon temperature of the Universe today. 

Equations~(\ref{equ:Y12eqx}) were integrated for different sets of masses of
shadow fermions varying between 30 and 300 GeV. 
The QZD coupling constant, $\alpha_Z \left({E}\right)$, values at energies $ \Lambda_Z <
E < 10^{23} \text{ GeV}$ are given for different sets of $m_1$ and $m_2$ in
Ref.~\cite{Hung2006a}. Within
the mass range we perform our relic density calculations, $\alpha_Z$ varies so
slowly and continuously that it can be obtained for any set of $m_1$ and $m_2$
by simple interpolation and extrapolation of the values provided in
Ref.~\cite{Hung2006a}. In this work, we have taken $\alpha_Z$ dependence
on $m_1$, $m_2$, and $E$ into account in our relic density calculations.
Nevertheless, it is worth mentioning that for a fixed $m_2$ and at a given $E$,
$\alpha_Z$ does not vary much as $m_1$ changes. For example, from Figs.~(1-3) of
Ref.~\cite{Hung2006a}, for $m_2 =$ 100 GeV and at $E =$ 150 GeV one obtains
\begin{align}
&\alpha_Z = 1.87500 \times 10^{-1}\qquad \text{or} \qquad  \alpha_Z^2 = 3.51563
\times 10^{-2} \quad \text{for} \quad m_1 = 1 \text{ GeV}  , \\
&\alpha_Z = 1.87149 \times 10^{-1}\qquad \text{or} \qquad  \alpha_Z^2 = 3.50247
\times 10^{-2} \quad \text{for} \quad m_1 = 10 \text{ GeV}  , \\
&\alpha_Z = 1.86567 \times 10^{-1}\qquad \text{or} \qquad  \alpha_Z^2 = 3.48074
\times 10^{-2} \quad \text{for} \quad m_1 = 50 \text{ GeV}  ,
\end{align}
which demonstrate how $\alpha_Z$ varies for $1{\text{ GeV}} \leqslant m_1 
\leqslant 50{\text{ GeV}}$
. The $\alpha_Z$ variation within such range (and similar $m_1$ ranges) is even
less noticeable in relic density calculations, since we are dealing with
$\alpha_Z^2$ in the annihilation cross sections. Our calculations showed that
one could safely use an average $\alpha_Z^2$ value over a wide range of $m_1$ values
without any sensible loss of accuracy. For instance, an $\alpha_Z^2 = 3.49961
\times 10^{-2}$ for the above range works just fine.

Ref.~\cite{Hung2006a} carries out RG analysis of QZD's coupling constant considering a
messenger field mass scale (mass of $\tilde{\bm{ \varphi}} _1^{\left( Z
\right)}$ the lighter messenger field) $m_{\varphi_1} = 300$ GeV and higher,
which points to when the QZD plasma decouples from the rest of the Universe. For our relic
density calculations, we always chose $m_{\varphi_1} > m_2$. It turns out that
the relic density of shadow fermions does not depend on the choice of $m_{\varphi_1}
>m_2$, as long as they are \textit{sufficiently} apart\footnote{The thermal contact between the shadow and visible sectors may still be in effect through virtual exchange of a messenger boson for some temperatures below the mass of the lighter messenger field. With $m_{\varphi _1 } $  being sufficiently larger than $m_2 $, the QZD plasma is ensured to have decoupled from the rest of the Universe before $\psi _2^{\left( Z \right)}$ enters its nonrelativistic epoch, decoupling from an \textit{isolated} QZD matter.}. That is mainly because the relic
densities of shadow fermions (or more generally WIMP's) are mostly determined in
their nonrelativistic epoch, i.e., for our case when $T_Z \le m_2$.

The decay of $\psi_2^{\left({Z}\right)}$ into a pair of SM leptons and
$\psi_1^{\left({Z}\right)}$ happens through a messenger field (see
Fig.~\ref{fig:decdiag} of Appendix~\ref{app:dec}). When the mass difference
$\Delta m = m_2 -m_1$ is not very large, the decay rate for one of the possible
decays can be given in an approximate form
\begin{equation}\label{equ:drate}
\Gamma_{21} \approx \frac{\alpha_ {\varphi_1}^2}{288 \pi}
\frac{m_2^5}{\left({m_2^2
-m_{\varphi_1}^2 }\right)^2 + m_{\varphi_1}^2 \Gamma_{\varphi_1}^2} \left({1- 8x
+8x^3 - x^4
-12x^2 \ln x}\right) \;,
\end{equation}
where $\alpha _ {\varphi_1} = g^2_{\varphi_1}/4 \pi$, $\Gamma_{\varphi_1}$ is
the
decay width of the messenger field and $x = m_1^2/m_2^2$. As already said, we
concentrate on the messenger field being sufficiently heavier than
$\psi_2^{\left({Z}\right)}$ where the ``singularity'' in the decay rate is not present, which can be seen
from the approximate from of $\Gamma_{21}$, 
Eq.~(\ref{equ:drate}). We shall explain the interesting case of $m_2 =
m_{\varphi_1}$ when we present our results in the next section. It is worth
mentioning, nevertheless, that such mass degeneracy poses no computational
difficulty due to the presence of the messenger field's decay width
$\Gamma_{\varphi_1}$.

On the other hand, $\alpha _ {\varphi_1}$ is constrained for the model to predict the observed baryon asymmetry through an initial lepton asymmetry produced in the decay of messenger fields~\cite{Hung2006a}. That requirement sets $\alpha _ {\varphi_1} \approx 2.9 \times 10^{-17}$, which will
consequently correspond to a long lifetime for $\psi_2^{\left({Z}\right)}$ (not less than
$10^7$ sec). For that reason, the decay rate of $\psi_2^{\left({Z}\right)}$ does
not effectively enter the relic density calculations\footnote{That means the
decay of $\psi_2^{\left({Z}\right)}$ is not determinant of the freeze-out
temperatures.},  where the evolution equations are dominated by the annihilation
processes. The remnant $\psi_2^{\left({Z}\right)}$'s (after the freeze-out)
decay into SM leptons and $\psi_1^{\left({Z}\right)}$'s anyway and we end up
with no relic for $\psi_2^{\left({Z}\right)}$ if the shadow fermion masses are
not degenerate. 
\section{Results}
The relic density of shadow fermions depend on two parameters: their masses,
$m_1$, and $m_2$. The masses affect the annihilation cross sections and
consequently the dynamics of the evolution equations. Our relic density
calculation results, therefore, are displayed either in terms of masses or mass
difference.

\begin{figure*}
\includegraphics{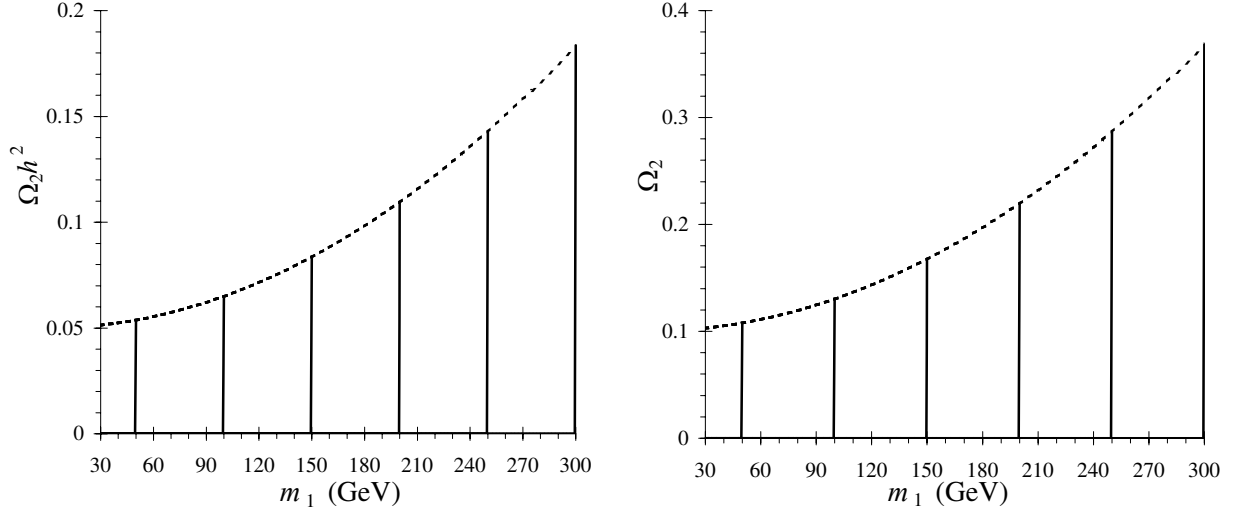}
\caption{\label{fig:o2}The relic density of $\psi_2^{\left({Z}\right)}$ versus
$\psi_1^{\left({Z}\right)}$'s mass $m_1$ at fixed $\psi_2^{\left({Z}\right)}$
masses: solid lines, two shadow fermions at $m_2$'s (from left to right) = 50,
100, 150, 200, 250, 300 GeV; dashed line, one shadow fermion. Note that 
$h = 0.732$ in this figure and throughout this work. For the dashed
line the horizontal axis is $m$.}
\end{figure*}

\begin{figure*}
\includegraphics{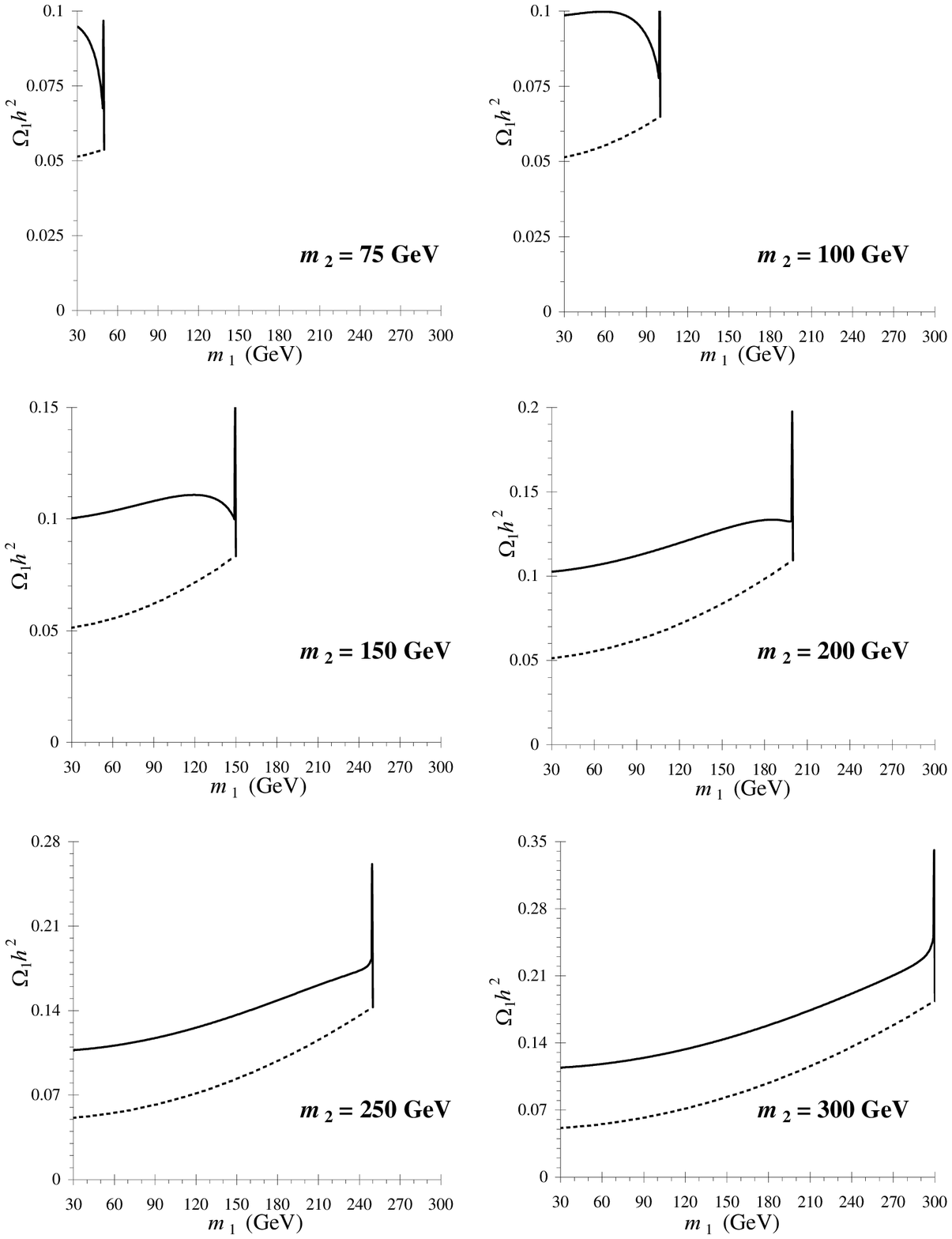}
\caption{\label{fig:o1h2}The relic density, $\Omega_1h^2$, of
$\psi_1^{\left({Z}\right)}$ versus $\psi_1^{\left({Z}\right)}$'s mass $m_1$ at
fixed $\psi_2^{\left({Z}\right)}$ masses: solid line, two shadow fermions;
dashed line, one shadow fermion. For the dashed line the horizontal axis is
$m$.}
\end{figure*}

\begin{figure*}
\includegraphics{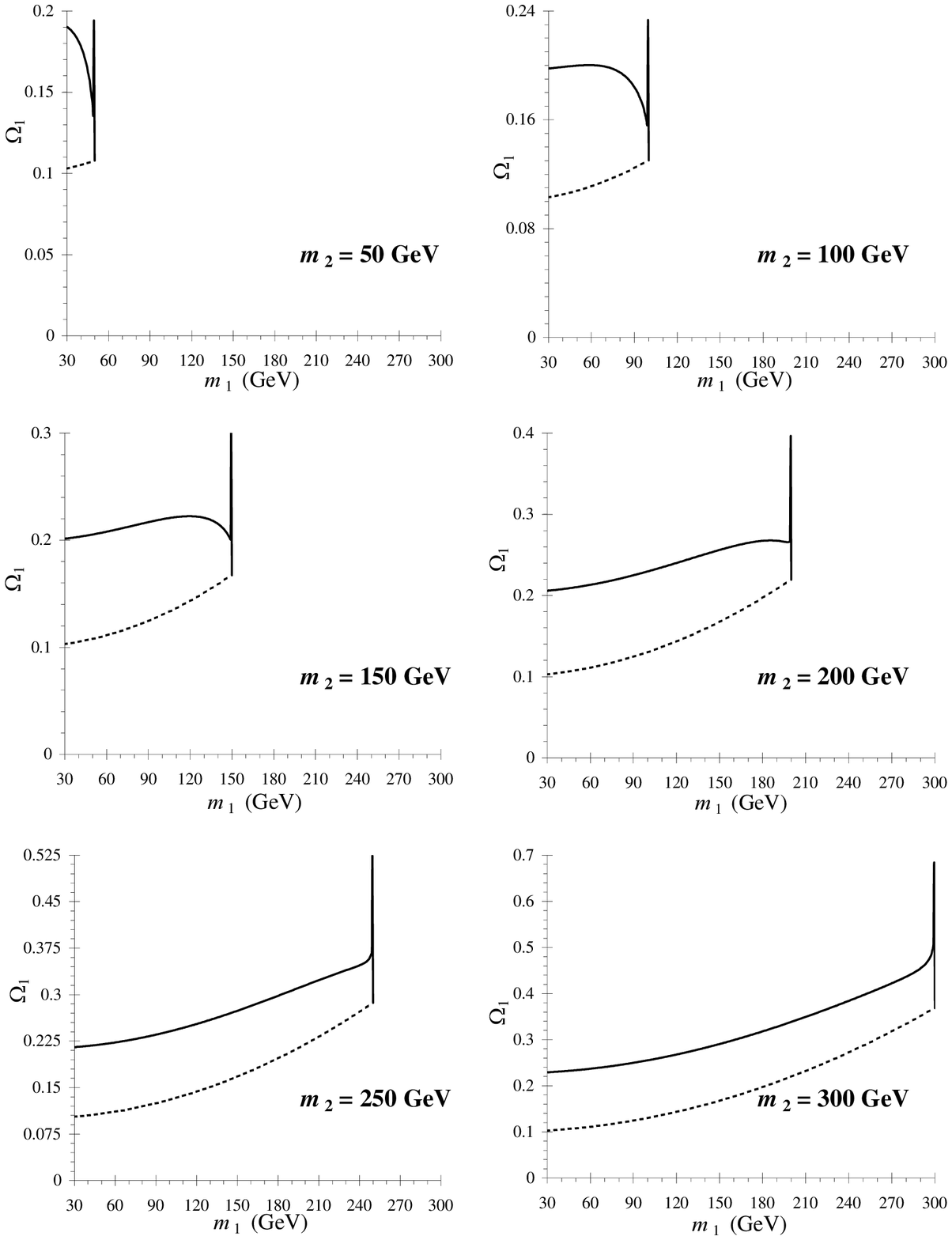}
\caption{\label{fig:o1}The relic density, $\Omega_1$, of
$\psi_1^{\left({Z}\right)}$ versus $\psi_1^{\left({Z}\right)}$'s mass $m_1$ at
fixed $\psi_2^{\left({Z}\right)}$ masses: solid line, two shadow fermions;
dashed line, one shadow fermion. For the dashed line the horizontal axis is
$m$.}
\end{figure*}

Suppose there were only one shadow fermion; in that case, the corresponding evolution equation
would be administered by shadow fermion's annihilation process and the expansion of the
Universe. Since the annihilation cross section into shadow gluons
and its thermal average $\left\langle {\sigma v} \right\rangle $
are inversely proportional to the mass squared, a heavier shadow fermion would
freeze out earlier than a lighter one, as it could not sustain a rate larger
than the Hubble rate for as long. That would allow less time (at
temperatures below the mass of the sole shadow fermion) for the Boltzmann factor
to diminish the density, which would result in a higher relic density compared
to a light shadow fermion's. This can be seen from the behavior shown by the
dashed line in Figs.~\ref{fig:o2}, \ref{fig:o1h2}, and \ref{fig:o1}, which
describes the density of $\psi_1^{\left({Z}\right)}$ or
$\psi_2^{\left({Z}\right)}$ if they were the sole fermion in the QZD particle
content. From those graphs, one sees that a heavy sole shadow fermion would have
a higher relic than a light one.

With two shadow fermions, however, there are two mechanisms governing the
evolution equations, besides the effect of the expanding Universe. There are
those reactions, which exhaust the phase space from the species and those that
populate it. The evolution of the number densities is determined by the
competition of those mechanisms. The outcome of such a competition, on the other
hand, depends on the masses of shadow fermions.

Of those mechanisms, the decay of $\psi_2^{\left({Z}\right)}$ plays no role in
the early dynamics of evolution equations. Briefly, that is because the lifetime
of $\psi_2^{\left({Z}\right)}$, which depends on $m_1$, $m_2$, and
$m_{\varphi_1}$,
turns out either too long or too short to be a factor in the determination of
freeze-out temperatures. For a well-separated set of $m_2$, and $m_{\varphi_1}$,
the lifetime of $\psi_2^{\left({Z}\right)}$ is within $10^7 {\text{ sec}} \lesssim
\tau _2  \lesssim 10^{13} {\text{ sec}}$ when $m_1
\ne m_2$, i.e, a nondegenerate case. That roughly corresponds to a temperature $1{\text{ keV}} \lesssim T
\lesssim 1{\text{ eV}}$, which is well after a typical freeze-out for
$\psi_2^{\left({Z}\right)}$. That means, the remainder of
$\psi_2^{\left({Z}\right)}$ will decay into $\psi_1^{\left({Z}\right)}$ and SM
leptons after the freeze-out, which leaves no present day abundance for
$\psi_2^{\left({Z}\right)}$. The decay of an unstable shadow fermion at such low
temperature into SM leptons can potentially disturb the cosmic microwave
background (CMB). That, as we shall see, will place a bound on the mass of
$\psi_2^{\left({Z}\right)}$ which determines the density of
$\psi_2^{\left({Z}\right)}$ at the time of its decay. With a mass degeneracy,
i.e., $m_1 = m_2$, of course $\psi_2^{\left({Z}\right)}$ is stable and decay is
irrelevant. In that case, since the annihilation channel into another is also
closed, we end up with two one-species cases: one for
$\psi_1^{\left({Z}\right)}$ and one for $\psi_2^{\left({Z}\right)}$.

When $m_2 = m_{\varphi_1}$, the decay width of the messenger field determines
the lifetime of $\psi_2^{\left({Z}\right)}$. As discussed in Ref.~\cite{Hung2006a}, the requirement for the lightest messenger field to decouple before decaying yields $\Gamma _{\varphi_1}   \approx m_{\varphi_1}  \alpha _{\varphi_1}  $, which is less than the expansion rate of the Universe, at $T=m_{\varphi_1}$. Since $\alpha _{\varphi_1}$ is of the order $\sim 10^{-17}$, a lifetime of $10^{ -
25} {\text{ sec}} \lesssim \tau _2  \lesssim 10^{ - 15} {\text{ sec}}$ for
$\psi_2^{\left({Z}\right)}$ is obtained. With such short lifetime,
$\psi_2^{\left({Z}\right)}$ decays well prior to the decoupling of QZD matter,
i.e., when QZD and the SM plasmas are in equilibrium. Effectively, that means we
are down to the one-species case, regardless of the value of $m_2$.

Thus, the annihilation processes and their competition will mainly decide for
the early dynamics of the evolution equations. At temperatures above the mass of
the heavier shadow fermion $\psi_2^{\left({Z}\right)} $,  both shadow fermions
contribute to the population of another through the annihilation process $\psi
_i^{\left( Z \right)} \bar \psi _i^{\left( Z \right)}  \to \psi _j^{\left( Z
\right)} \bar \psi _j^{\left( Z \right)} $. As temperature decreases, the
contribution of the lighter shadow fermion  $\psi_1^{\left({Z}\right)} $ into
the population of $\psi_2^{\left({Z}\right)} $ diminishes until it stops at an
energy when it is no longer kinematically allowed. From there on,
$\psi_2^{\left({Z}\right)}$ will lose pairs monotonically due to its
annihilations into shadow gluons and $\psi_1^{\left({Z}\right)} $ pairs, while
$\psi_1^{\left({Z}\right)}$ receives pairs from $\psi_2^{\left({Z}\right)} $'s
annihilation and at the same time loses pairs due to annihilation into shadow
gluons.

The annihilation into $\psi_1^{\left({Z}\right)}$ provides an additional channel
for $\psi_2^{\left({Z}\right)}$ to keep up with the expansion rate of the
Universe and therefore delay the freeze-out. This reduces the density of
$\psi_2^{\left({Z}\right)}$ prior to its freeze-out, compared to the one-species
case, in two ways: (i) $\psi_2^{\left({Z}\right)}$ pairs are lost into
$\psi_1^{\left({Z}\right)}$ pairs in addition to those lost into shadow gluons,
(ii) the Boltzmann factor for temperatures $T_Z <m_2$ can act on
$\psi_2^{\left({Z}\right)}$'s density for a longer time.

All this, though, depends on how apart $\psi_1^{\left({Z}\right)}$ and
$\psi_2^{\left({Z}\right)}$ are, masswise, at a fixed $m_2$. Since the available
phase space for $\psi _2^{\left( Z \right)} \bar \psi _2^{\left( Z \right)}  \to
\psi _1^{\left( Z \right)} \bar \psi _1^{\left( Z \right)}$ increases with the
mass difference $\Delta m = m_2 - m_1$, we expect $\psi_2^{\left({Z}\right)}$'s
density at freeze-out becoming small for an increasing $\Delta m$ due to a
growing annihilation rate. On the other hand, a small mass difference reduces
the phase space for the annihilation process and therefore increases the
density. Knowing this is important in understanding the constraint on
$\psi_2^{\left({Z}\right)}$'s density at the time of decay. Since the remaining
$\psi_2^{\left({Z}\right)}$'s will decay anyway, there will be no relic for
$\psi_2^{\left({Z}\right)}$ if $m_1 \ne m_2$, which is reflective in
Fig.~\ref{fig:o2}, where $\psi_2^{\left({Z}\right)}$'s relic densities are
displayed in solid lines for different $m_2$'s as $m_1$ varies. The relic
density of $\psi_2^{\left({Z}\right)}$ falls down rapidly when the mass
difference between the two shadow fermions is enough to allow the decay before
our time and therefore to deplete the phase space from
$\psi_2^{\left({Z}\right)}$ pairs. The maximum relic density, however, is always
at $m_1 = m_2$, where the annihilation cross section, $\sigma_{ij}$, and the
decay rate $\Gamma_{21}$ are vanishing and it is essentially the one-species
case.

The situation for $\psi_1^{\left({Z}\right)}$ is more complicated. The relic
density of $\psi_1^{\left({Z}\right)}$ is shown through a solid line in
Figs.~\ref{fig:o1h2}, and \ref{fig:o1} for different $m_2$'s as $m_1$ varies.
For an extremely heavy $\psi_1^{\left({Z}\right)}$, i.e. $m_1  = m_2$,
$\psi_1^{\left({Z}\right)}$'s relic density coincides with the one-species case,
as expected. As $\Delta m$ deviates from zero $\psi_2^{\left({Z}\right)}$ starts
to dispense $\psi_1^{\left({Z}\right)}$ pairs into the phase space (by
annihilation earlier, and decay later) and thus $\Omega_1$ increases. Prior to
freeze-out, this positive contribution comes from the pair annihilation of
$\psi_2^{\left({Z}\right)}$ into $\psi_1^{\left({Z}\right)}$, which will face a
growing competition from $\psi_1^{\left({Z}\right)}$'s annihilation channel into
shadow gluons, as $m_1$ declines. Since the annihilation cross section into
shadow gluons grows for small masses, it will start to contend the rate
of the extra $\psi_1^{\left({Z}\right)}$ pairs coming from
$\psi_2^{\left({Z}\right)}$'s annihilation. For that reason, as $m_1$ decreases,
the annihilation channel into shadow gluons depletes the phase space from
$\psi_1^{\left({Z}\right)}$ pairs more effectively and therefore
$\psi_1^{\left({Z}\right)}$'s density before the freeze-out, which consequently diminishes its
relic $\Omega_1$. After the freeze-out, the remnant of
$\psi_2^{\left({Z}\right)}$ will decay into $\psi_1^{\left({Z}\right)}$ and
lifts $\Omega_1$, very much by a constant, except at small $\Delta m$'s where
$\psi_2^{\left({Z}\right)}$'s density is larger.
\begin{figure*}
\includegraphics{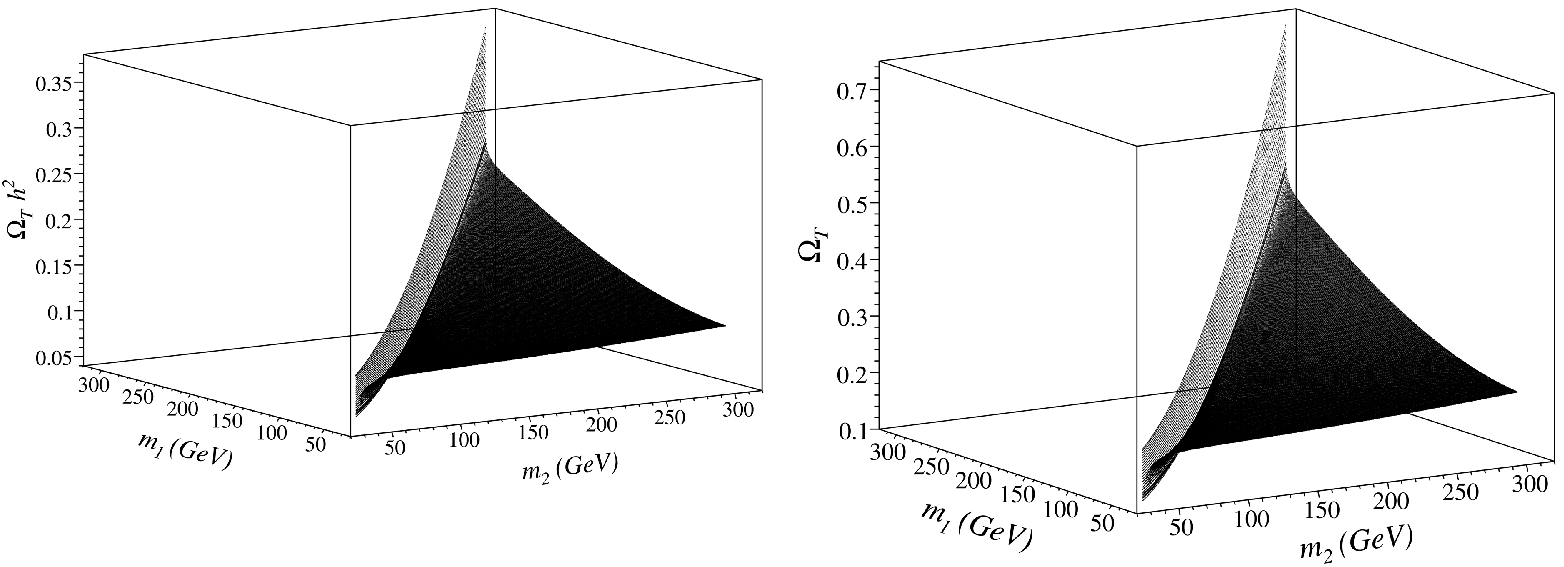}
\caption{\label{fig:oT3d}The three-dimensional depiction of the total relic
density of shadow fermions as both masses vary.}
\end{figure*}

\begin{figure*}
\includegraphics{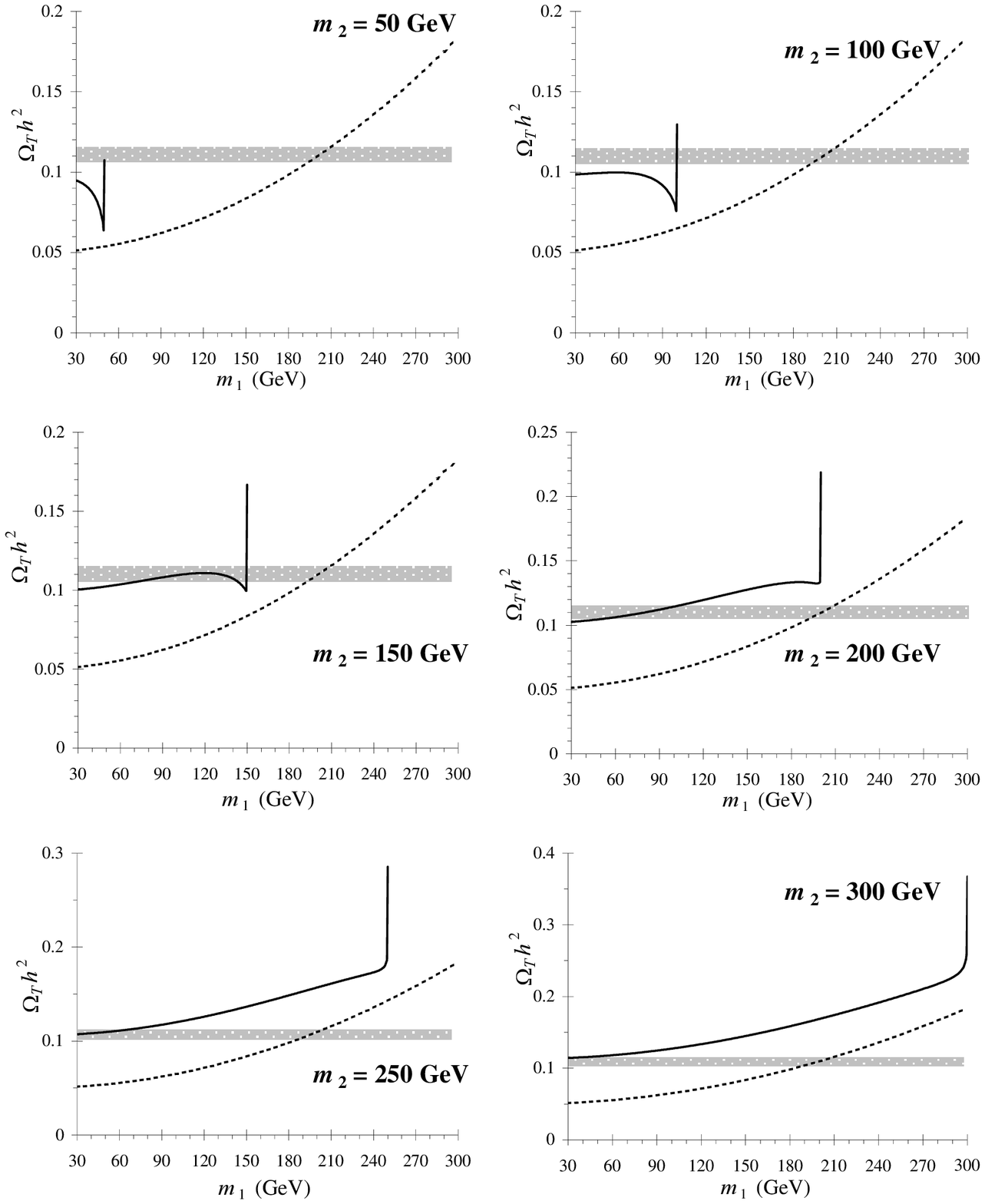}
\caption{\label{fig:oTh2}The total relic density of shadow fermions, $\Omega_T
h^2$, versus $\psi_1^{\left({Z}\right)}$'s mass $m_1$ at fixed
$\psi_2^{\left({Z}\right)}$ masses: solid line, two shadow fermions; dashed
line, one shadow fermion. For the dashed line, the horizontal axis is the mass
of the sole shadow fermion. The gray band represents the allowed density from
WMAP3 and all data sets~\cite{WMAP3}.}
\end{figure*}

\begin{figure*}
\includegraphics{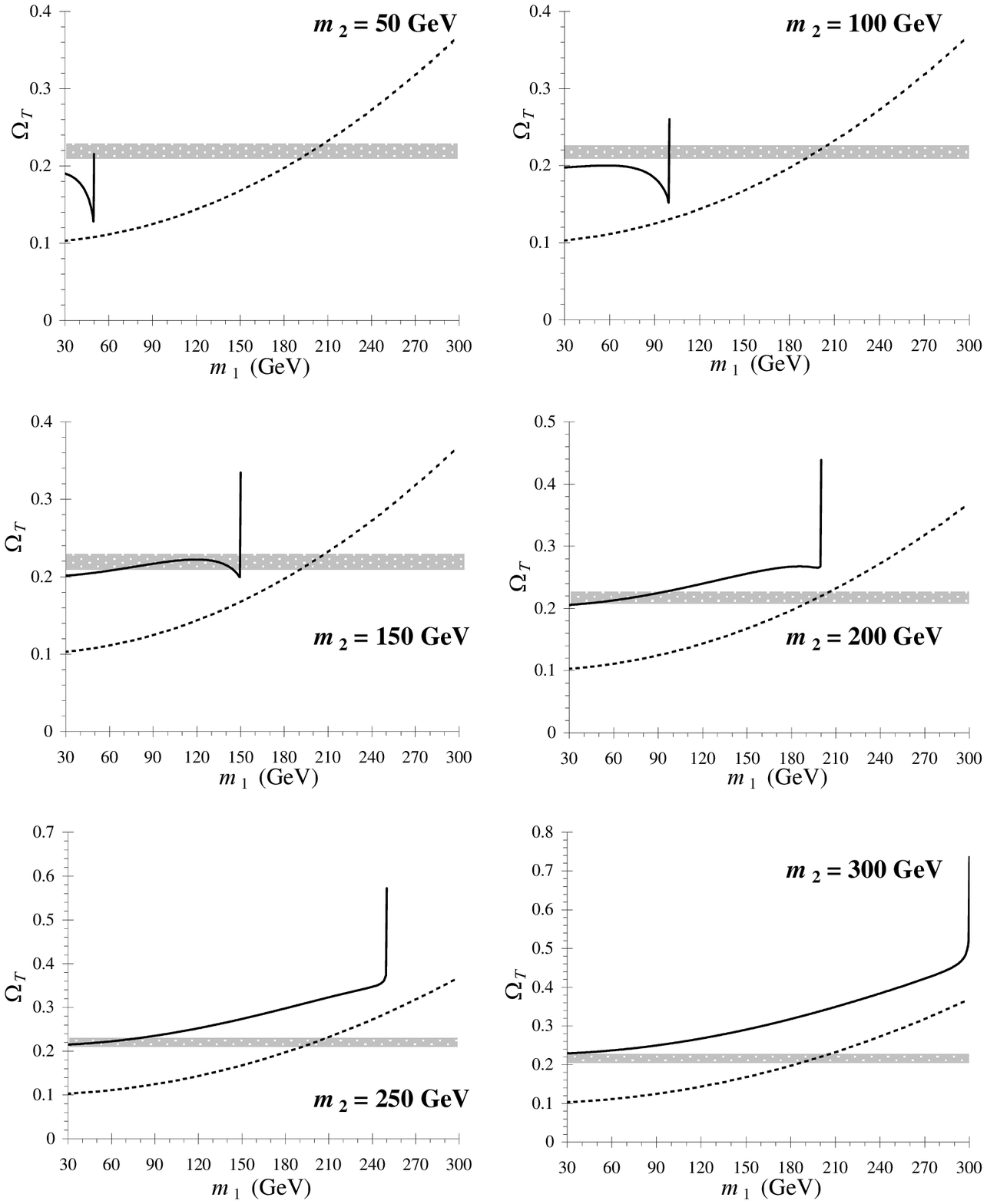}
\caption{\label{fig:oT}The total relic density of shadow fermions, $\Omega_T$,
versus $\psi_1^{\left({Z}\right)}$'s mass $m_1$ at fixed
$\psi_2^{\left({Z}\right)}$ masses: solid line, two shadow fermions; dashed
line, one shadow fermion. For the dashed line, the horizontal axis is the mass
of the sole shadow fermion. The gray band represents the allowed density from
WMAP3 and all data sets~\cite{WMAP3}.}
\end{figure*}

For a nondegenerate mass case, $\psi_1^{\left({Z}\right)}$'s relic density is
what remains of shadow fermions. It is only at $m_1 = m_2$ that the relic
consists of both shadow fermions (equally so). To be inclusive of the degenerate
case, the total relic density of shadow fermions $\Omega_T = \Omega_1 +
\Omega_2$ is presented in Fig.~ \ref{fig:oT3d} against both masses and in
Figs.~\ref{fig:oTh2}, \ref{fig:oT} against $m_1$ at fixed $m_2$'s, where the
one-species case is also presented. The gray areas in Figs.~\ref{fig:oTh2}, and
\ref{fig:oT} indicate the current bounds on the dark matter density from WMAP3
and all data sets~\cite{WMAP3}.

It can be seen in Figs.~\ref{fig:oTh2}, and \ref{fig:oT} that the total relic
density $\Omega_T$ increases as $m_1$ does, attaining a sharp maximum for the
degenerate case, as if there were two ``one-species" shadow fermions. On the
other hand, $\Omega_T$ also increases with $m_2$, which means for staying in the
cosmologically allowed region a larger and larger mass difference would be
needed. The two extremes are at $m_2 = 50$ GeV, where the degenerate case is
just making it to the allowed region, and $m_2 = 300$ GeV, where a large mass
difference is needed to stay relevant.
\begin{figure*}
\includegraphics{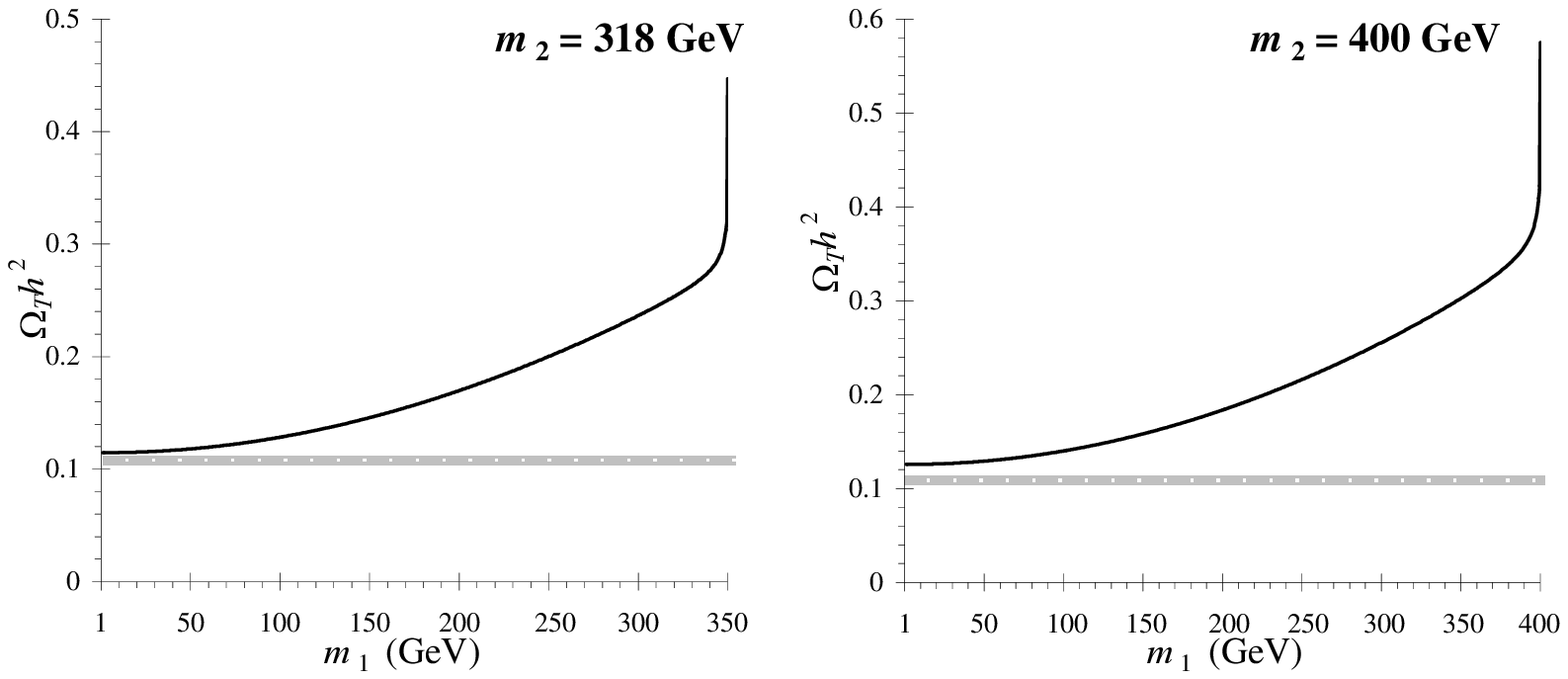}
\caption{\label{fig:ohigh}The total relic density of shadow fermions versus
$\psi_1^{\left({Z}\right)}$'s mass $m_1$ at high $\psi_2^{\left({Z}\right)}$
masses. The gray band represents the allowed density from WMAP3 and all data
sets~\cite{WMAP3}.}
\end{figure*}

\begin{figure*}
\includegraphics{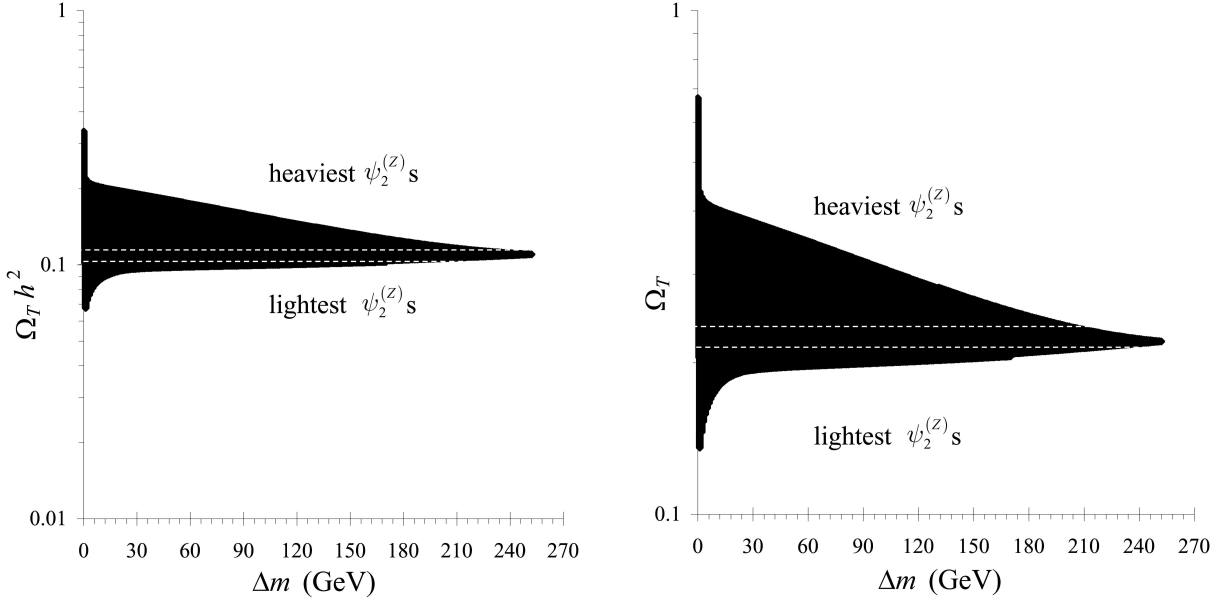}
\caption{\label{fig:ovdm}The total relic density of shadow fermions versus their
mass difference $\Delta m$. The white dashed lines indicate the bounds from
WMAP3 and all data sets \cite{WMAP3}.}
\end{figure*}

This can also be seen by looking at Fig.~\ref{fig:ovdm}, in which the total
relic density is displayed versus $\Delta m$ and the bounds are shown with two
white dashed lines. We conclude that for $m_2 < 50$ GeV, the total relic density
is not enough to account for the total dark matter, even though the shadow
fermions would still be relic particles taking on a fraction of the dark matter
in the Universe.

On the other hand, for $m_2 \gtrapprox 320$ GeV, the total density of shadow
fermions go beyond the upper bound and give unacceptable values even if we
extend the mass difference to an extreme where $m_1 = 1$ GeV.  That is shown in
Fig.~\ref{fig:ohigh}, where the total relic density at $m_2 = 318$ GeV is only
viable for a large mass difference of about 317 GeV and at $m_2 = 400$ GeV is
no longer relevant. By going to such an extreme mass difference, we place a
\textit{naive} bound on the mass of the heavier shadow fermion, i.e., $m_2
\approxeq 320$ GeV, above which the total relic density is no longer viable. 

There are, however, more restrictive bounds on $m_2$ coming from the decay of
$\psi_2^{\left({Z}\right)}$ into SM leptons at low temperature, and its potential
disturbance of the CMB of the Universe. We demand that
\begin{enumerate}
  \item The density of $\psi_2^{\left({Z}\right)}$ at the time of decay could not exceed that of the SM particles.
  \item The CMB density disturbance caused by the late decay of $\psi_2^{\left({Z}\right)}$ would not violate the CMB fluctuation, which has been observed to be at $10^{-5}$ level~\cite{WMAP3}.
\end{enumerate}

\begin{figure*}
\includegraphics{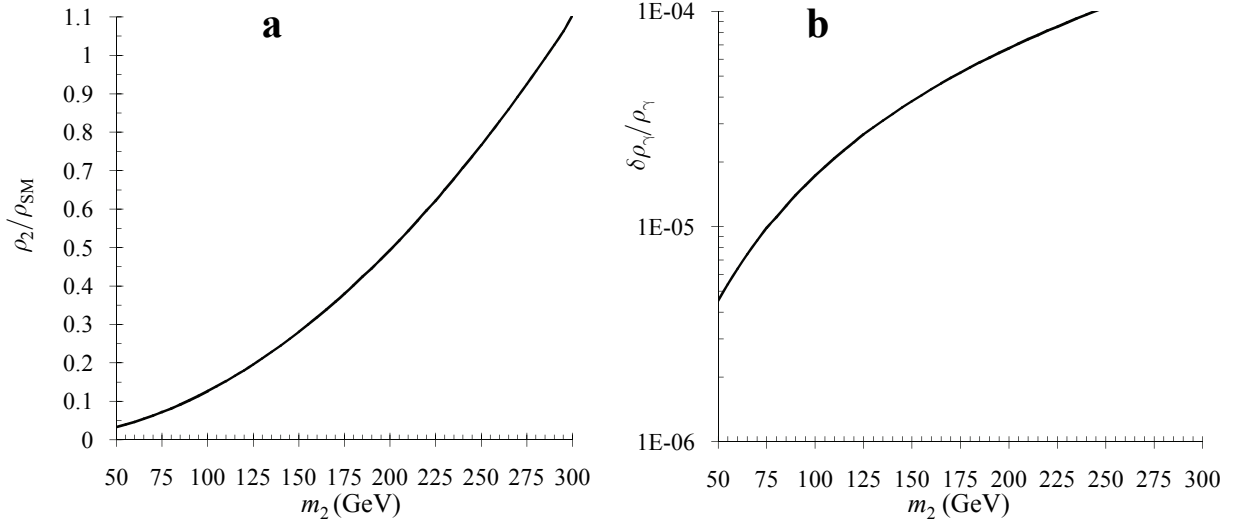}
\caption{\label{fig:cons} Cosmological constraints on the mass of the heavier shadow fermion: a) The ratio of the density of
$\psi_2^{\left({Z}\right)}$ to the density of the SM matter, right before it
starts to decay, versus the mass of $\psi_2^{\left({Z}\right)}$; b) The amplitude of CMB density disturbance from the late decay of $\psi_2^{\left({Z}\right)}$ versus the mass of $\psi_2^{\left({Z}\right)}$ .}
\end{figure*}

Figure~\ref{fig:cons} illustrates these two conditions in graphs versus the mass of
$\psi_2^{\left({Z}\right)}$. In Fig.~\ref{fig:cons}~a, ${{\rho _2 } \mathord{\left/ {\vphantom
{{\rho _2 } {\rho _{{\text{SM}}} }}} \right. \kern-\nulldelimiterspace} {\rho
_{{\text{SM}}} }}$, i.e., the density of $\psi_2^{\left({Z}\right)}$ to the
density of the SM matter -- right before the decay --  is plotted, which shows that the density of $\psi_2^{\left({Z}\right)}$ remains less than that of the SM particles for $m_2 \leqslant 285$ GeV. The possible CMB density disturbance, ${{\delta \rho _\gamma  } \mathord{\left/ {\vphantom {{\delta \rho _\gamma  } {\rho _\gamma  }}} \right. \kern-\nulldelimiterspace} {\rho _\gamma  }}$, that the late decay of $\psi _2^{\left( Z \right)}$ can create is shown in Fig.~\ref{fig:cons}~b. The CMB density disturbance goes above the $10^{-5}$ order for $\psi _2^{\left( Z \right)}$s heavier than  245 GeV.
The two above conditions, therefore, place a strong bound of 245 GeV on $\psi _2^{\left( Z \right)}$'s mass.

As we discussed, the lifetime of $\psi_2^{\left({Z}\right)}$ could be very short
if $\psi_2^{\left({Z}\right)}$ and the messenger field were degenerate in mass.
In that case, the total relic density of shadow fermions is simply that of the
one-species case and it yields the right density for masses between 190 and 210
GeV.
 
For $50 \text{ GeV} \lessapprox m_2 \lessapprox 245\text{ GeV}$, the total relic density of
shadow
fermions can account for the amount of the dark matter in the Universe,
depending on the mass difference. The total relic density lies within the
observational bounds with small and even zero mass difference for light
$\psi_2^{\left({Z}\right)}$'s and with large mass differences when
$\psi_2^{\left({Z}\right)}$ is heavy.

\section{Summary}
We solved evolution equations for number densities of shadow fermions and
obtained their total present-day density. The heavier shadow fermion turned out
to be long lived if its mass differs from that of the messenger field. In that case,
our results revealed an upper bound on the mass of the heavier shadow fermion, i.e.,
$m_2 \approx 245$ GeV, above which its \textit{late} decay can potentially disturb the CMB density of the Universe beyond the measured fluctuation level of $10^{-5}$.

For lighter shadow fermions, the total relic density can account for the entire
dark matter of the Universe depending on the mass combination of shadow
fermions. When the total density falls short of the observationally suggested
density, it still, for most of masses, provides significant fraction of the dark
matter of the Universe.

Our results showed that if the heavier shadow fermion's mass is large,
considerable mass differences would be needed to comply with experimental
bounds. On the other hand, if the heavier shadow fermion's mass is small, little
or even no mass differences suffice to give the right relic density. In that
sense, degenerate and near-degenerate mass cases become relevant at low mass
scales, but not for less than 50 GeV. 

A very short lifetime is expected for the heavier shadow fermion if its mass is
the same as that of the messenger field. In that case, the calculations reduce
to a one-species case. Our results suggest that a sole shadow fermion must have a
mass of about 190~--~210 GeV to account for the whole dark matter of the
Universe. 

Last but not least, possible detections of the shadow fermion CDM candidates are
briefly discussed in Ref.~\cite{Hung2006a}. Needless to say, more work along
this line is warranted for this model.
\begin{acknowledgments}
This work was supported, in part, by the U.S. Department of Energy under grant
No. DE-A505-89ER40518. One of us (PQH) would like to thank Goran Senjanovic,
Alexei Smirnov and ICTP for the hospitality where this manuscript was finished.
\end{acknowledgments}

\appendix
\section{Annihilation cross sections}\label{app:ACS}
\begin{figure*}
\setlength{\unitlength}{1.0cm}
\begin{picture}(15,5)(0.8,0)
\put(0.17,4.33){$k,a,\lambda$}
\put(3.27,4.33){$k',b,\lambda'$}
\put(13,4.33){$p_j,n,s_j$}
\put(15.4,4.33){$p'_j,n',s'_j$}
\put(0.16,0.42){$p,l,s$}
\put(3.19,0.42){$p',l',s'$}
\put(13,0.42){$p_i,l,s_i$}
\put(15.4,0.42){$p'_i,l',s'_i$}
\put(15.1,2.4){$q$}
\put(10.35,2.6){$q_s$}
\put(6.2,2.15){$q_u$}
\put(2.05,2.15){$q_t$}
\put(4.05,-0.5){$\psi_i^{\left({Z}\right)} \bar{\psi}_i^{\left({Z}\right)}
\rightarrow \bm{A}^{\left({Z}\right)}\bm{A}^{\left({Z}\right)}$}
\put(13.15,-0.5){$\psi_i^{\left({Z}\right)} \bar{\psi}_i^{\left({Z}\right)}
\rightarrow \psi_j^{\left({Z}\right)} \bar{\psi}_j^{\left({Z}\right)}$}
\includegraphics{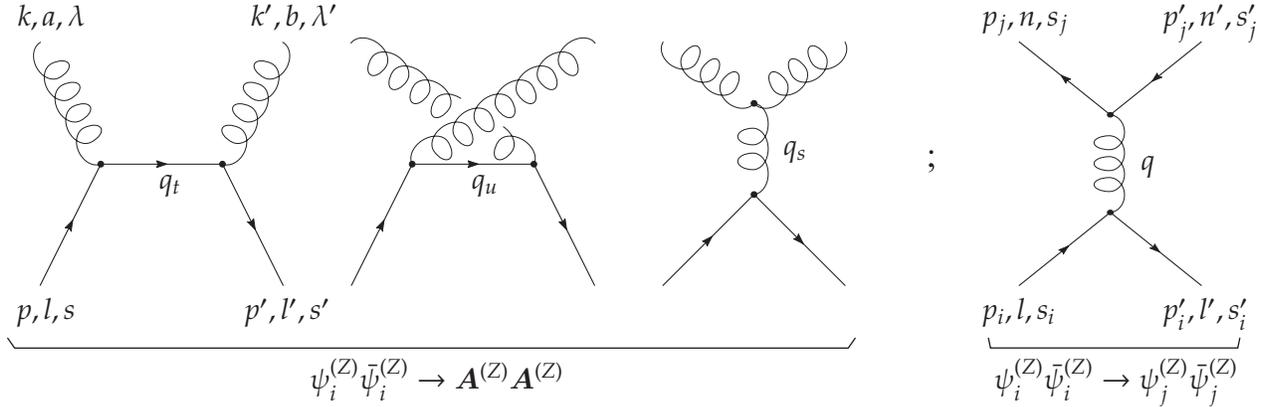}
\end{picture}
\caption{\label{fig:anndiags} Tree-level diagrams for: left, pair annihilation
of shadow fermions into shadow gluons; right, pair annihilation of one type of
shadow fermions into a pair of another type.}
 \end{figure*}

The pair annihilation of shadow fermions can yield either two shadow gluons
 or another  pair of shadow fermions. The
diagrams, to leading order, for both processes are displayed in
Fig.~\ref{fig:anndiags}, where the former process happens through three diagrams
in $t$, $u$, and $s$ channels and the latter in $s$ channel. In those diagrams,
$p$, $p'$, $k$, $k'$, $p_i$, $p'_i$, $p_j$, $p'_j$ are momenta, $l$, $l'$, $n$,
$n'$ and $a,b$ are the QZD colors of shadow fermions and shadow gluons, $s$,
$s'$, $s_i$, $s'_i$, $s_j$, $s'_j$ and $\lambda, \lambda'$ are the spins of
fermions and final polarizations of shadow gluons, and $q_t$, $q_u$, $q_s$, $q$
are momentum transfers.
\subsection{Annihilation into two shadow gluons}
We first compute the total annihilation cross section for a pair of shadow
fermions into two shadow gluons denoted by three diagrams in
Fig.~\ref{fig:anndiags}. We carry out the computation for a fermion triplet with
mass $m$, generically. The covariant amplitude $\mathcal{M}$ of the diagrams
simply reads 
\begin{align}
\mathcal{M} =&  -g_Z^2  \left( {\hat T^b} \right) _{l'n} \left( {\hat T^a}
\right)_{nl} \bar v^{s'}_{l'} \left( {p'} \right) \slashed{\epsilon} _b^{*
\lambda '} \left( {k'} \right)\frac{\slashed{k}-\slashed{p}+m }{{q_t^2 - m^2 }}
\slashed{\epsilon} _a^{* \lambda } \left( k \right)u^{s}_l \left( {p} \right)
\cr 
  &-g_Z^2  \left( {\hat T^a } \right)_{l'n} \left( {\hat T^b } \right)_{nl} \bar
v^{s'}_{l'} \left( {p'} \right) \slashed{\epsilon} _a^{*\lambda } \left( k
\right)\frac{\slashed{k'}-\slashed{p}+m }
{{q_u^2 - m^2 }} \slashed{\epsilon} _b^{*\lambda '} \left( {k'} \right)u^{s}_l
\left( {p} \right) \cr
&- ig_Z^2 \varepsilon^{abc} \left( {\hat T^c } \right)_{l'l}  \bar v^{s'}_{l'}
\left( {p'} \right)  \frac{\gamma ^\sigma  \epsilon _a^{*\mu \lambda } \left( k
\right) \epsilon _b^{*\nu \lambda '} \left( {k'} \right) }{{q_s^2 }}
\Big[\left({k-k'}\right)_\sigma \eta_{\mu \nu} \cr
  &+ \left({q_s-k'}\right)_\mu \eta_{\nu \sigma} + \left({k-q_s}\right)_\nu
\eta_{\sigma \mu} \Big] u^{s}_l \left( {p} \right) ,
\end{align}
where, e.g., $\epsilon _a^{\mu \lambda } \left( k \right)$ is the shadow gluon
polarization four-vector, with $\lambda$ indicating its polarization state,
$q_t=k-p$, $q_u=k-p'$, and $q_s=k+k'=p+p'$. We are looking for an unpolarized
cross section with the initial degrees of freedom averaged over and the final
ones summed over, which corresponds to the averaged squared amplitude
\begin{equation}\label{equ:Mav}
\overline {\left| \mathcal{M} \right|^2 }  = \sum\limits_{\lambda ,\lambda '}
{\frac{1}
{4}\sum\limits_{s,s'} {\sum\limits_{a,b} {\frac{1}
{9}\sum\limits_{l,l',n} {\left| \mathcal{M} \right|^2  } } } } .
\end{equation}
We may even further compactify $\mathcal{M}$ in the form
\begin{equation}
 \mathcal{M} = \epsilon _a^{*\mu \lambda} \left( {k}\right) \epsilon _b^{* \nu
\lambda '} \left( {k'}\right) \mathcal{K}^{ab}_{\mu \nu}\, ,
\end{equation}
with
\begin{align}
\mathcal{K}^{ab}_{\mu \nu}=& -g_Z^2 \bar v^{s'}_{l'} \left( {p'} \right)
\Bigg[\hat T ^b_{l'n} \hat T ^a_{nl} \gamma_\nu\frac{\slashed{k}-\slashed{p}+m
}{{t - m^2 }} \gamma_\mu +\hat T ^a_{l'n} \hat T ^b_{nl} \gamma_\mu
\frac{\slashed{k'}-\slashed{p}+m }{{u - m^2 }} \gamma_\nu \cr
&+i\hat T ^c_{l'l} \frac{\gamma ^\sigma  \varepsilon _{abc}}{{s }}
\left[{\left({k-k'}\right)_\sigma \eta_{\mu \nu}+ \left({q_s-k'}\right)_\mu
\eta_{\nu \sigma} + \left({k-q_s}\right)_\nu \eta_{\sigma \mu} } \right] \Bigg]
u^{s}_l \left( {p} \right) ,
\end{align}
where $s$, $t$, and $u$ are the Mandelstam variables of the process. Therefore,
$\left|{\mathcal{M}}\right|^2$ will have a compact form
\begin{equation}\label{equ:MS}
\left|{\mathcal{M}}\right|^2 =\epsilon _{a'}^{\alpha \lambda} \left( {k}\right)
\epsilon _a^{*\mu \lambda} \left( {k}\right) \epsilon _{b'}^{ \beta \lambda '}
\left( {k'}\right) \epsilon _b^{* \nu \lambda '} \left( {k'}\right)
\mathcal{K}^{*a'b'}_{\alpha \beta} \mathcal{K}^{ab}_{\mu \nu}\;.
\end{equation}
The sums over the QZD colors of the squared amplitude, in Eq.~(\ref{equ:Mav}),
result in five types of traces, namely
\begin{equation}
\sum\limits_{a,b} {\mathrm{Tr}\left( {\hat T^a \hat T^b \hat T^a \hat T^b }
\right)}  = 6 \,,
\end{equation}
\begin{equation}
\sum\limits_{\scriptstyle a,b,c,d}  {\varepsilon _{acd} \varepsilon _{bcd}
\mathrm{Tr} \left( {\hat T^a \hat T^b } \right)}  = 12 \,,
\end{equation}
\begin{equation}\label{equ:ts}
\sum\limits_{a,b,c} { i\varepsilon _{abc} \mathrm{Tr} \left( {\hat T^b \hat T^a
\hat T^c } \right)}  = 6 \,,
\end{equation}
\begin{equation}\label{equ:us}
\sum\limits_{a,b,c} { i\varepsilon _{abc} \mathrm{Tr} \left( {\hat T^a \hat T^b
\hat T^c } \right)}  = - 6 \, ,
\end{equation}
\begin{equation}
\sum\limits_{a,b} {\mathrm{Tr} \left( {\hat T^a \hat T^b \hat T^b \hat T^a }
\right)}  = 12 \, ,
\end{equation}
knowing which yields
\begin{align}\label{equ:M21}
\frac{1}{9} \sum\limits_{\text{colors}} { \left|{\mathcal{M}}\right|^2} =
\frac{1}{9} \Big[ & 6\left|{\mathfrak{M}_t}\right|^2 + 6
\left|{\mathfrak{M}_u}\right|^2 +12  \left|{\mathfrak{M}_s}\right|^2 \cr
&+ 12 \times 2\text{Re}\left({ \mathfrak{M}_t^*\mathfrak{M}_u}\right) -6 \times
2\text{Re}\left({ \mathfrak{M}_u^*\mathfrak{M}_s}\right) +6 \times
2\text{Re}\left({ \mathfrak{M}_s^*\mathfrak{M}_t}\right)\Big],
\end{align}
where the amplitudes $\mathfrak{M}_t$, $\mathfrak{M}_u$, $\mathfrak{M}_s$ are
colorless, having only the Lorentz degrees of freedom. Evaluating $\overline
{\left| \mathcal{M} \right|^2 }$ also includes summations over initial spins and
final polarizations. The sum over spins is simply the familiar $\gamma$-matrix
manipulation. On the other hand, sum over final polarizations involves terms
like
\[
\sum\limits_\lambda  {\epsilon ^{\alpha \lambda } \left( k \right)\epsilon
^{*\mu \lambda } \left( k \right)} \quad\quad \text{and} \quad\quad
\sum\limits_{\lambda'}  {\epsilon ^{\beta \lambda' } \left( k' \right) \epsilon
^{*\nu \lambda' } \left( k'	 \right)} .
\]
To avoid closed loop diagrams containing ghost lines, we use the covariant form
\begin{equation}
\sum\limits_\lambda  {\epsilon ^{\mu \lambda } \left( k \right)\epsilon ^{*\nu
\lambda } \left( k \right)}  =  - \eta ^{\mu \nu }  + 2\frac{{k^\mu  k'^\nu   +
k^\nu  k'^\mu  }}
{s},
\end{equation}
which preserves the gauge invariance and has the same effect as
\[
\sum\limits_\lambda  {\epsilon ^{\mu \lambda } \left( k \right)\epsilon ^{*\nu
\lambda } \left( k \right)}  =  - \eta ^{\mu \nu }  + {\text{ghost terms}} \,.
\]

Considering all that, the spin averaged and polarization summed
$\mathfrak{M}$-terms of Eq.~(\ref{equ:M21}), in terms of the Mandelstam
variables of the process are 
\begin{subequations}\label{equ:Ms}
\begin{equation}
\sum\limits_{\mathrm{polarizations}}\!\!\!\!
{\frac{1}{4}\sum\limits_{\mathrm{spins}} {\left| {\mathfrak{M}_t } \right|^2}} 
= g_Z^4 \left[ {\frac{{2\left( {u - m^2 } \right)}}
{{t - m^2 }} - \frac{{4m^2 }}
{{t - m^2 }} - \frac{{8m^4 }}
{{\left( {t - m^2 } \right)^2 }}} \right],
\end{equation}
\begin{equation}
\sum\limits_{\mathrm{polarizations}}\!\!\!\!
{\frac{1}{4}\sum\limits_{\mathrm{spins}} {\left| {\mathfrak{M}_u } \right|^2}} 
= g_Z^4 \left[ {\frac{{2\left( {t - m^2 } \right)}}
{{u - m^2 }} - \frac{{4m^2 }}
{{u - m^2 }} - \frac{{8m^4 }}
{{\left( {u - m^2 } \right)^2 }}} \right],
\end{equation}
\begin{equation}
\sum\limits_{\mathrm{polarizations}}\!\!\!\!
{\frac{1}{4}\sum\limits_{\mathrm{spins}} {\left| {\mathfrak{M}_s } \right|^2}} 
= \frac{{4g_Z^4 }}
{{s^2 }}\left[ {m^2 \left( {2u - s} \right) - m^4  - s^2  - u\left( {u + s}
\right)} \right],
\end{equation}
\begin{equation}
\sum\limits_{\mathrm{polarizations}}\!\!\!\!
{\frac{1}{4}\sum\limits_{\mathrm{spins}} {2\mathrm{Re}\left({\mathfrak{M}_t^*
\mathfrak{M}_u}\right)}}  = \frac{{-4m^2g_Z^4 }}
{{\left( {t - m^2 } \right)\left( {u - m^2 } \right)}}\left[ {4m^2  + \left( {t
- m^2 } \right)+\left( {u - m^2 } \right) } \right],
\end{equation}
\begin{equation}
\sum\limits_{\mathrm{polarizations}}\!\!\!\!
{\frac{1}{4}\sum\limits_{\mathrm{spins}} {2\mathrm{Re}\left({\mathfrak{M}_u^*
\mathfrak{M}_s}\right)}}  = \frac{{4g_Z^4 }}
{{s\left( {u - m^2 } \right)}}\left[ {m^4  + m^2 \left( {s - 2u} \right) + u^2 }
\right],
\end{equation}
\begin{equation}
\sum\limits_{\mathrm{polarizations}}\!\!\!\!
{\frac{1}{4}\sum\limits_{\mathrm{spins}} {2\mathrm{Re}\left({\mathfrak{M}_s^*
\mathfrak{M}_t}\right)}}  = \frac{{4g_Z^4 }}
{{s\left( {t - m^2 } \right)}}\left[ {m^4  + m^2 \left( {u - t} \right) - \left(
{u + s} \right)^2 } \right].
\end{equation}
\end{subequations}
And finally, in terms of momenta, the unpolarized amplitude squared is given by
\begin{align}\label{equ:M22} \nonumber
  \overline {\left|{\mathcal{M}} \right|^2 }  =& \frac{1}
{9}\Bigg\{6g_Z^4 \left( { - 2\frac{{m^4 }}
{{\left( {p \cdot k} \right)^2 }} + 2\frac{{p \cdot k'}}
{{p \cdot k}} + 2\frac{{m^2 }}
{{p \cdot k}}} \right) + 6g_Z^4 \left( { - 2\frac{{m^4 }}
{{\left( {p \cdot k'} \right)^2 }} + 2\frac{{p \cdot k}}
{{p \cdot k'}} + 2\frac{{m^2 }}
{{p \cdot k'}}} \right) \cr 
   &+ 12\frac{{4g_Z^4 }}
{{\left( {p + p'} \right)^4 }}\Big[ m^4  + 4m^2 p \cdot k - 3m^2 \left( {p + p'}
\right)^2  - \left( {p + p'} \right)^4 \cr
 &- \left( {m^2  - 2p \cdot k'} \right)\left( {m^2  + 2p \cdot k} \right) \Big] 
   + 12g_Z^4 \left( { - 4\frac{{m^4 }}
{{\left( {p \cdot k} \right)\left( {p \cdot k'} \right)}} + 2\frac{{m^2 }}
{{p \cdot k'}} + 2\frac{{m^2 }}
{{p \cdot k}}} \right)\cr
& - 6\frac{{ - 4g_Z^4 }}
{{\left( {p + p'} \right)^2 }}\left( {2m^2  + m^2 \frac{{p \cdot k}}
{{p \cdot k'}} + 2p \cdot k'} \right)  \cr
  &+ 6\frac{{ - 4g_Z^4 }}
{{\left( {p + p'} \right)^2 }}\left( { - 2m^2  + m^2 \frac{{p \cdot k'}}
{{p \cdot k}} - 2p \cdot k} \right) \Bigg\}.
\end{align}
The differential cross section in the center-of-mass (CM) frame, where $p =
\left( {E,{\bf{p}}} \right)$ and $p' = \left( {E, - {\bf{p}}} \right)$,  reads
\begin{align}
  \left({\frac{{d\sigma }}
{{d\cos \theta }}}\right)_{\mathrm{CM}} =& \frac{{\pi \alpha _Z^2 }}
{{3E^2 }}\frac{1}
{v}\left[ {\frac{{1 + v^2 \cos ^2 \theta }}
{{1 - v^2 \cos ^2 \theta }} - \left( {1 - v^2 } \right)\frac{{1 + v^2 \cos ^2
\theta }}
{{\left( {1 - v^2 \cos ^2 \theta } \right)^2 }}} \right. \cr 
  & \left. { + 2\frac{{1 - v^4 }}
{{1 - v^2 \cos ^2 \theta }} - \left( {1 - v^2 } \right)\frac{{v\cos \theta }}
{{1 - v\cos \theta }} - \frac{{v^2 }}
{2}\cos ^2 \theta  + v^2  - \frac{3}
{2}} \right],
\end{align}
where $v = {\left|{\bf{p}}\right| \mathord{\left/ {\vphantom
{\left|{\bf{p}}\right| E}} \right.
\kern-\nulldelimiterspace} E}$ is the velocity of annihilating particles in the
CM frame and $\alpha _Z  = {{g_Z^2 } \mathord{\left/ {\vphantom {{g_Z^2 } {4\pi
}}} \right. \kern-\nulldelimiterspace} {4\pi }}$. The total cross section
then follows as
\begin{equation}\label{equ:rel1}
\sigma_{\mathrm{CM}}  = \frac{{\pi \alpha _Z^2 }}
{{3E^2 }}\frac{1}
{v}\left[ {\frac{2 - v^4 }
{v}\ln \left( {\frac{{1 + v}}
{{1 - v}}} \right) + \frac{{1 - v^2 }}
{v}\ln \left( {1 - v} \right) - \frac{{v^2 }}
{6} - \frac{5}
{2}} \right].
\end{equation}
In non-relativistic limit when $E \to m$ and $v \ll 1$, we obtain, neglecting
$\mathcal{O}\left({v^2}\right)$,
\begin{equation}\label{equ:nr1}
\sigma_{\mathrm{CM}} ^{{\mathrm{nr}}}  = \frac{{\pi \alpha _Z^2 }}
{{3m^2 }}\left( {\frac{1}
{{2v}} - \frac{{10v}}
{3} - \frac{1}
{2}} \right).
\end{equation}

The relativistic cross section in the lab frame (the rest frame of one of the
annihilating particles) can be obtained as well. In terms of the velocity of the
incoming particle in the lab frame $v$, it is
\begin{align}\label{equ:lab1}
  \sigma_{\mathrm{Lab}}  = \frac{{2\pi \alpha _Z^2 }}
{{3m^2 }}&\frac{{\sqrt {1 - v^2 }  - \left( {1 - v^2 } \right)}}
{{v^2 }} \cr
&\left[ {\frac{{v^4  + 8v^2  + 4\left( {2 - v^2 } \right)\sqrt {1 - v^2 }  - 8}}
{{2v^2  - v^4  - 2v^2 \sqrt {1 - v^2 } }}\ln \left( {\frac{{1 + v  - \sqrt {1 -
v^2 } }}
{{v  - 1 - \sqrt {1 - v^2 } }}} \right)} \right. \cr 
  & \;\;\,+ 2\frac{{\sqrt {1 - v^2 }  - \left( {1 - v^2 } \right)}}
{{2 - v^2  - 2\sqrt {1 - v^2 } }}\ln \left( {\frac{{1 + v  - \sqrt {1 - v^2 } }}
{{v }}} \right) \cr
&\;\;\, \left. {- \frac{{1 - \sqrt {1 - v^2 } }}
{{6v }} - \frac{{5v }}
{{2 - 2\sqrt {1 - v^2 } }}} \right].
\end{align}
\subsection{Annihilation of shadow fermions into each other}
The annihilation of shadow fermions into each other can occur through a shadow
gluon, or the scalar field $\phi_Z$. The smallness of the Yukawa coupling of
$\phi_Z$ field, nonetheless, makes its channel rather negligible compared to the
shadow gluon channel. For that reason and to leading order, the annihilation of
a pair of $\psi_i^{(Z)}$ with mass $m_i$ into a pair of $\psi_j^{(Z)}$ with mass
$m_j$ is considered through the corresponding diagram of
Fig.~\ref{fig:anndiags}. The covariant amplitude $\mathcal{M}$ of the diagram
reads 
\[
  \mathcal{M} = g_Z^2 \left( {\hat T^b } \right)_{nn'}\left( {\hat T^a }
\right)_{l'l} \bar u^{s_j}_n \left( {p_j} \right) \gamma _\mu v^{s'_j}_{n'}
\left( {p'_j} \right) \frac{{\delta _{ab}}}
{{q^2}} \bar v^{s'_i}_{l'} \left( {p'_i} \right) \gamma ^\mu  u^{s_i}_l\left(
{p_i} \right),
\]
where, $q^2=s=\left({p_j+p'_j}\right)^2=\left({p_i+p'_i}\right)^2$. Once again,
we are looking for an unpolarized cross section involving
\[
\overline {\left| \mathcal{M} \right|^2 }  = \frac{1}
{4} \sum\limits_{s_i,s'_i} {\sum\limits_{s_j,s'_j} {\frac{1}
{9}\sum\limits_{a,b} {\sum\limits_{l,l'} {\sum\limits_{n,n'} {\left| \mathcal{M}
\right|^2 } } } } } .
\]
The gauge algebra calculations, which contain sums over QZD colors of the
squared amplitude, result in a trace of the form
\begin{equation}
\sum\limits_{\scriptstyle a,b}  { \mathrm{Tr} \left( {\hat T^a \hat T^b }
\right) \mathrm{Tr} \left( {\hat T^b \hat T^a } \right)} = 12.
\end{equation}
After summing over QZD colors, we obtain
\begin{equation}\label{equ:Mbarp}
\frac{1}{9} \sum\limits_{\text{colors}} { \left|{\mathcal{M}}\right|^2} =
\frac{1}{9} \times 12 \times \left| \mathfrak{M} \right|^2 ,
\end{equation}
where the amplitude $\mathfrak{M}$ is colorless and only has Lorentz degrees of
freedom. The Lorentz algebra including summations over initial and final spins
yields
\begin{align}
\overline {\left| \mathfrak{M} \right|^2 } &= \frac{1}
{4} \sum\limits_{\mathrm{spins}} {\left| \mathfrak{M} \right|^2} \cr
&= \frac{2g_Z^4}
{s^2}\left[{ \left({t -m_j^2-m_i^2}\right)^2 +  \left({u -m_j^2-m_i^2}\right)^2
+ 2 \left({m_j^2+m_i^2}\right)s} \right].
\end{align}
where $s$, $t$, and $u$ are the Mandelstam variables of the process. The
unpolarized squared amplitude is then given by
\begin{equation}\label{equ:MbarMand}
 \overline {\left| \mathcal{M} \right|^2 } = \frac{8g_Z^4}
{3s^2}\left[{ \left({t -m_j^2-m_i^2}\right)^2 +  \left({u -m_j^2-m_i^2}\right)^2
+ 2 \left({m_j^2+m_i^2}\right)s} \right],
\end{equation}
and in terms of momenta
\begin{equation}\label{equ:Mbars}
 \overline {\left| \mathcal{M} \right|^2 } = \frac{32g_Z^4}
{3s^2}\left[{\left({p'_j \cdot p'_i}\right)\left({p_j \cdot p_i}\right) +
\left({p'_j \cdot p_i}\right) \left({p_j \cdot p'_i}\right) +m_j^2 p_i  \cdot
p'_i + m_i^2 p_j  \cdot p'_j + 2m_j^2m_i^2} \right].
\end{equation}
In the CM frame, where $ p_i=(E,\bf{p})$ and $p'_i=(E,-\bf{p})$, the
differential cross section is
\begin{align}\label{equ:DCSij}
\left({\frac{{d\sigma }}
{{d\cos \theta }}}\right)_{\mathrm{CM}} = \frac{{\pi \alpha _Z^2 }}
{3}\frac{1}
{{4m_i^2 }}\frac{{1 - v_i^2 }}
{v_i}\sqrt {1 - \frac{{m_j^2 }}
{{m_i^2 }}\left( {1 - v_i^2 } \right)} &\left[ {2 - v_i^2 \left( {1 -
\frac{{m_j^2 }}
{{m_i^2 }}\left( {1 - v_i^2 } \right)} \right)\cos ^2 \theta  }\right.\cr
&\;\;\left.{+ 2\left( {1 + \frac{{m_j^2 }}
{{m_i^2 }}} \right)\left( {1 - v_i^2 } \right)} \right].
\end{align}
where $v_i = {\left|{\bf{p}}\right| \mathord{\left/ {\vphantom
{\left|{\bf{p}}\right| E}} \right.
\kern-\nulldelimiterspace} E}$ is the velocity of the annihilating particles
(i.e., $\psi_i^{\mathrm{Z}} \bar{\psi}_i^{\mathrm{Z}}$) in the CM frame. The
total cross section is then obtained as
\begin{align}\label{equ:relsigma2}
\sigma_{\mathrm{CM}}  = &\frac{{\pi \alpha _Z^2 }}
{3}\frac{1}
{{m_i^2 }}\frac{{1 - v_i^2 }}
{v_i}\sqrt {1 - \frac{{m_j^2 }}
{{m_i^2 }}\left( {1 - v_i^2 } \right)}\cr
& \left[ {1 - v_i^2  - \frac{{v_i^2 }}
{6}\left( {1 - \frac{{m_j^2 }}
{{m_i^2 }}\left( {1 - v_i^2 } \right)} \right) + \left( {1 + \frac{{m_j^2 }}
{{m_i^2 }}\left( {1 - v_i^2 } \right)} \right)} \right].
\end{align}
The nonrelativistic limit of the total cross section, when $v_i \ll 1$, and $E
\to m_i $, can be easily obtained, neglecting $\mathcal{O}\left({v_i^2}\right)$,
\begin{equation}
\sigma_{\mathrm{CM}} ^{{\mathrm{nr}}}  = \frac{{\pi \alpha _Z^2 }}
{3}\frac{{\sqrt {1 - {{m_j^2 } \mathord{\left/
 {\vphantom {{m_j^2 } {m_i^2 }}} \right.
 \kern-\nulldelimiterspace} {m_i^2 }}} }}
{{m_i^2 }}\left( {\frac{{2 + {{m_j^2 } \mathord{\left/
 {\vphantom {{m_j^2 } {m_i^2 }}} \right.
 \kern-\nulldelimiterspace} {m_i^2 }}}}
{v_i} - \frac{{7 - {{m_j^2 } \mathord{\left/
 {\vphantom {{m_j^2 } {m_i^2 }}} \right.
 \kern-\nulldelimiterspace} {m_i^2 }}}}
{6}v_i} \right). 
\end{equation}
In the lab frame, the relativistic total cross section can be also given as
\begin{align}
  \sigma_{\mathrm{Lab}}  = &\frac{{2\pi \alpha _Z^2 }}
{{3m_i^2 }}\frac{{v_i^2  + \sqrt {1 - v_i^2 }  - 1}}
{{v_i \left( {1 - \sqrt {1 - v_i^2 } } \right)}}\left( {1 - 2\frac{{m_j^2 }}
{{m_i^2 }}\frac{{v_i^2  + \sqrt {1 - v_i^2 }  - 1}}
{{v_i^2 }}} \right)^{{1 \mathord{\left/
 {\vphantom {1 2}} \right.
 \kern-\nulldelimiterspace} 2}}  \cr 
  &\left[ {1 - \frac{{2\left( {1 - \sqrt {1 - v_i^2 } } \right) - v_i^2 }}
{{6v_i^2 }}\left( {1 - 2\frac{{m_j^2 }}
{{m_i^2 }}\frac{{v_i^2  + \sqrt {1 - v_i^2 }  - 1}}
{{v_i^2 }}} \right)} \right. \cr 
  &\;\;\left. { + 2\left( {1 + \frac{{m_j^2 }}
{{m_i^2 }}} \right)\frac{{v_i^2  + \sqrt {1 - v_i^2 }  - 1}}
{{v_i^2 }}} \right],
\end{align}
where $v_i$ here is the velocity of the incoming particle (i.e., beam) in the
lab frame.
\section{The heavier shadow fermion's decay}\label{app:dec}
\begin{figure*}
\setlength{\unitlength}{1.0cm}
\begin{picture}(15,5)(-6,0)
\put(-0.18,4.05){$p,r$}
\put(1.4,4.05){$p',r'$}
\put(1.8,1.5){$k$}
\put(3,4.05){$p_1,l, s_1$}
\put(0.3,-0.4){$p_2,j, s_2$}
\includegraphics{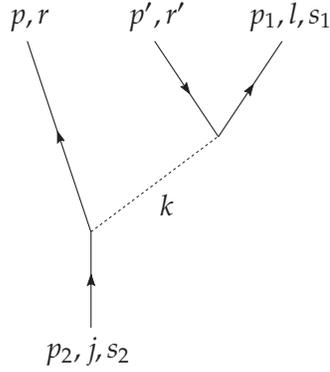}
\end{picture}
\caption{\label{fig:decdiag} The decay of $\psi_2^{\left({Z}\right)}$ into SM
leptons and $\psi_1^{\left({Z}\right)}$ through a scalar messenger field.}
 \end{figure*}
The decay of $\psi _2^{\left( Z \right)}\to l \bar{l'} \psi _1^{\left( Z
\right)}$ is possible through the lighter messenger field $\tilde{\bm{ \varphi}}
_1^{\left( Z \right)}$ (either real or virtual,
depending on masses) and can yield any pair of SM leptons. Due to considerably
small leptonic masses, when compared to shadow fermions', we carry out the decay
rate calculation in the limit of massless SM leptons. In that sense, the 
decay rate for $\psi _2^{\left( Z \right)}$ through $\tilde{\bm{ \varphi}}
_1^{\left( Z \right)}$ with mass
$m_{\varphi_1}$ and a Yukawa coupling $g_{\varphi_1}$, representing any of
$g_{\tilde
\varphi _1 m}^i$, can be computed. The process, to
leading order, occurs through the diagram of Fig.~\ref{fig:decdiag}. The
covariant amplitude $\mathcal{M}$ of the diagram is
\begin{equation}
\mathcal{M} = -g^2_{\varphi_1}  \bar u^{r} \left( {p} \right)_L  u^{s_2}_{j}
\left(
{p_2} \right)_R \frac{\delta _{jl}}{{k^2 - m_{\varphi_1}^2 + i m_{\varphi_1}
\Gamma_{\varphi_1} }} \bar u^{s_1}_{l} \left( {p_1} \right) _R v^{r'} \left(
{p'}
\right)_L,
\end{equation}
where $k = p_2 -p$, $\Gamma_{\varphi_1}$ is the decay width of the messenger
field
and momenta $p_1$, $p_2$ refer to those of shadow fermions, while $p$, $p'$ are
the momenta of the SM leptons. Similar to previous cases, we are looking for an
unpolarized decay rate with the initial degrees of freedom averaged over and the
final ones summed over, which corresponds to the averaged squared amplitude
\begin{equation}\label{equ:Mavd1}
\overline {\left| \mathcal{M} \right|^2 }  = \frac{1}
{2}\sum\limits_{s_1,s_2, r,r'} {\frac{1}
{3} \sum\limits_{j,l} {\left| \mathcal{M} \right|^2  } }  .
\end{equation}
There is not much of $\gamma$-matrix algebra involved in computing $\overline
{\left| \mathcal{M} \right|^2 }$, which easily gives
\begin{equation}\label{equ:Mavd2}
\overline {\left| \mathcal{M} \right|^2 }  = \frac{8g^4_{\varphi_1}}{3}
\frac{\left({p_1 \cdot p'}\right)\left({p_2 \cdot p}\right)}{|{k^2 -
m_{\varphi_1}^2 + i m_{\varphi_1}\Gamma_{\varphi_1} }|^2} \;.
\end{equation}

Finally, the decay rate, in the rest frame of  $\psi _2^{\left( Z \right)}$, can
be found after the usual three-body decay kinematical considerations, which
yields the integral form
\begin{equation}\label{equ:rate}
\Gamma _{\psi _2^{\left( Z \right)} }  = \frac{{\alpha _{\varphi_1} ^2 }}
{{72m_2^2 \pi }}\int_{\left({m_2 - m_1}\right)^2 }^0 {d\varpi ^2 \frac{{\sqrt
{\left( {m_1^2  + m_2^2  - \varpi ^2 } \right)^2  - 4m_1^2 m_2^2 } }}
{{m_2^2  - m_{\varphi_1} ^2  + m_{\varphi_1} ^2 \Gamma _{\varphi_1} ^2  - \varpi
^2 }}\left[
{\left( {m_2^2  - m_1^2 } \right)^2  + \left( {m_2^2  + m_1^2 } \right)^2 \varpi
^2  - 2\varpi ^4 } \right]} ,
\end{equation}
where $\alpha _{\varphi_1}   = {{g_{\varphi_1} ^2 } \mathord{\left/ {\vphantom
{{g_{\varphi_1} ^2 } {4\pi }}} \right. \kern-\nulldelimiterspace} {4\pi }}$ and
$\varpi ^2  = \left( {p_2  - p_1 } \right)^2$. The above decay rate behaves
according to an $m_2^5$ dependence for $m_2 < m_{\varphi_1} + m_1$,  and an
$m_2^3$
dependence for $m_2 > m_{\varphi_1} + m_1$.
\section{Thermal averaging}\label{app:TA}
The thermal averaging of $\sigma v$ (i.e., the annihilation cross section times
the relative velocity) is discussed in Ref.~\cite{Gondolo1991}, where a compact
single integral for $\left\langle {\sigma v} \right\rangle $ is provided. The
authors of Ref.~\cite{Gondolo1991} explain that the thermal averaging of
relativistic $\sigma v$ in the cosmic comoving frame and the lab frame are
equivalent but they differ from the $\left\langle {\sigma v} \right\rangle $
obtained in the CM frame. They stress that this difference is only significant
in the relativistic limit. To stay relativistically covariant they introduce
$\left\langle {\sigma v_{{\text{M\o{}l}}} } \right\rangle $, for the cosmic
comoving frame, where $v_{\text{M\o{}l}}$ is defined in terms of the velocities
of the two annihilating particles. The relation $\left\langle {\sigma
v_{{\text{M\o{}l}}} } \right\rangle  = \left\langle {\sigma v_{{\text{Lab}}} }
\right\rangle  \ne \left\langle {\sigma v_{{\text{CM}}} } \right\rangle $ holds,
in relativistic limit, anyway. To evaluate the thermal averages for our
annihilation processes, we make use of the relativistically-valid single
integral of Ref.~\cite{Gondolo1991}, which is
\begin{equation}\label{equ:av}
\left\langle {\sigma v_{{\text{M\o{}l}}} } \right\rangle  = \frac{{2x}}
{{K_2^2 \left( x \right)}}\int_0^\infty  {d\epsilon \sqrt \epsilon  \left( {1 +
2\epsilon } \right)K_1 \left( {2x\sqrt {1 + \epsilon } } \right)\sigma }
v_{{\text{lab}}} \; ,
\end{equation}
where $x = {m \mathord{\left/ {\vphantom {m T}} \right.
\kern-\nulldelimiterspace} T}$ ($m$ is the mass of the annihilating particles),
$K_i \left( x \right)$ is the modified Bessel function of order $i$ and 
\begin{equation}
\epsilon  = \frac{{s - 4m^2 }}
{{4m^2 }} \; ,
\end{equation}
\begin{equation}
 v_{\mathrm{lab}}  = \frac{{2\sqrt {\epsilon \left( {1 + \epsilon } \right)} }}
{{1 + 2\epsilon }} \; ,
\end{equation}
with  $s$ being the usual Mandelstam variable for the annihilation process. The
annihilation cross sections of shadow fermions are available analytically (see
Appendix~\ref{app:ACS}). Therefore, the thermal averages of interest can be
written with the help of Eq.~(\ref{equ:av}) in closed integral forms. The
integrals then can be evaluated numerically for given masses.

For the annihilation of a pair of shadow fermions into shadow gluons $\psi_i
^{\left( Z \right)} \bar \psi_i ^{\left( Z \right)}  \to
\bm{A}^{\left({Z}\right)}\bm{A}^{\left({Z}\right)}$, the
thermal average after simplification reads
\begin{align}
\left\langle {\sigma_{iA} v_{iA}} \right\rangle  = \frac{{4\pi
\alpha _Z^2 }}
{{3m_i^2 }}\frac{x_i}
{{K_2^2 \left( x_i \right)}} & \int_0^\infty  {d\epsilon K_1 \left( {2x_i\sqrt
{1 + \epsilon } } \right)  \left[ {\frac{{\epsilon ^2  + 4\epsilon  + 2}}
{{\left( {1 + \epsilon } \right)^{{3 \mathord{\left/
 {\vphantom {3 2}} \right.
 \kern-\nulldelimiterspace} 2}} }}\ln \left( {1 + \sqrt {\frac{\epsilon }
{{1 + \epsilon }}} } \right)} \right.}  \cr 
   & \;\;\;\;\;\;\; \left. {- \frac{{\epsilon ^2  + 3\epsilon  + 1}}
{{\left( {1 + \epsilon } \right)^{{3 \mathord{\left/
 {\vphantom {3 2}} \right.
 \kern-\nulldelimiterspace} 2}} }}\ln \left( {1 - \sqrt {\frac{\epsilon }
{{1 + \epsilon }}} } \right)
    - \frac{1}
{6}\frac{{\epsilon ^{{3 \mathord{\left/
 {\vphantom {3 2}} \right.
 \kern-\nulldelimiterspace} 2}} }}
{{1 + \epsilon }} - \frac{5}
{2}\sqrt \epsilon  } \right], 
\end{align}
where $x_i={m_i \mathord{\left/ {\vphantom {m_i T_Z}} \right.
\kern-\nulldelimiterspace} T_Z}$. For the annihilation of one pair of shadow
fermions into a pair of another, $\psi _i^{\left( Z \right)} \bar \psi
_i^{\left( Z \right)}  \to \psi _j^{\left( Z \right)} \bar \psi _j^{\left( Z
\right)} 
$, the corresponding thermal average is
\begin{equation}
 \left\langle {\sigma_{ij} v_{ij}} \right\rangle  = \frac{{4\pi \alpha _Z^2 }}
{{3m_i^2 }}\frac{x_i}
{{K_2^2 \left( x_i \right)}}\int_0^\infty  {d\epsilon K_1 \left( {2x_i\sqrt {1 +
\epsilon } } \right)\frac{{\sqrt \epsilon  }}
{{1 + \epsilon }}\sqrt {1 - \frac{{m_j^2 }}
{{m_i^2 }}\frac{1}
{{1 + \epsilon }}} \left[ {2 + \frac{{5\epsilon }}
{6} + \frac{{m_j^2 }}
{{m_i^2 }}\frac{{1 + {{7\epsilon } \mathord{\left/
 {\vphantom {{7\epsilon } 6}} \right.
 \kern-\nulldelimiterspace} 6}}}
{{1 + \epsilon }}} \right]}.
\end{equation}


\begin{thebibliography}{18}
\expandafter\ifx\csname natexlab\endcsname\relax\def\natexlab#1{#1}\fi
\expandafter\ifx\csname bibnamefont\endcsname\relax
  \def\bibnamefont#1{#1}\fi
\expandafter\ifx\csname bibfnamefont\endcsname\relax
  \def\bibfnamefont#1{#1}\fi
\expandafter\ifx\csname citenamefont\endcsname\relax
  \def\citenamefont#1{#1}\fi
\expandafter\ifx\csname url\endcsname\relax
  \def\url#1{\texttt{#1}}\fi
\expandafter\ifx\csname urlprefix\endcsname\relax\def\urlprefix{URL }\fi
\providecommand{\bibinfo}[2]{#2}
\providecommand{\eprint}[2][]{\url{#2}}

\bibitem[{\citenamefont{Hung}(2005)}]{Hung2005}
\bibinfo{author}{\bibfnamefont{P.~Q.} \bibnamefont{Hung}},
  \href{http://arxiv.org/abs/hep-ph/0504060}{
  [arXiv:hep-ph/0504060v2]}.

\bibitem[{\citenamefont{Hung}(2006{\natexlab{a}})}]{Hung2006a}
\bibinfo{author}{\bibfnamefont{P.~Q.} \bibnamefont{Hung}},
  \bibinfo{journal}{Nucl. Phys. B} \textbf{\bibinfo{volume}{747}},
  \bibinfo{pages}{55} (\bibinfo{year}{2006}{\natexlab{a}}),
  \href{http://arxiv.org/abs/hep-ph/0512282}{
  [arXiv:hep-ph/0512282v2]}.

\bibitem[{\citenamefont{Hung}(2007{\natexlab{a}})}]{Hung2007a}
\bibinfo{author}{\bibfnamefont{P.~Q.} \bibnamefont{Hung}},
  \href{http://arxiv.org/abs/0707.2791}{
  [arXiv:0707.2791v1]}.

\bibitem[{\citenamefont{Hung}(2007{\natexlab{b}})}]{Hung2007b}
\bibinfo{author}{\bibfnamefont{P.~Q.} \bibnamefont{Hung}}, \bibinfo{journal}{J.
  Phys. A} \textbf{\bibinfo{volume}{40}}, \bibinfo{pages}{6871}
  (\bibinfo{year}{2007}{\natexlab{b}}),
  \href{http://arxiv.org/abs/astro-ph/0612245v1}{
  [arXiv:astro-ph/0612245v1]}.

\bibitem[{\citenamefont{Hung and Mosconi}(2006)}]{Hung2006d}
\bibinfo{author}{\bibfnamefont{P.~Q.} \bibnamefont{Hung}} \bibnamefont{and}
  \bibinfo{author}{\bibfnamefont{P.}~\bibnamefont{Mosconi}},
  \href{http://arxiv.org/abs/hep-ph/0611001}{
  [arXiv:hep-ph/0611001v2]}.

\bibitem[{\citenamefont{Das and Laperashvili}(2007{\natexlab{a}})}]{Das}
\bibinfo{author}{\bibfnamefont{C.~R.} \bibnamefont{Das}} \bibnamefont{and}
  \bibinfo{author}{\bibfnamefont{L.~V.} \bibnamefont{Laperashvili}},
  \href{http://arxiv.org/abs/0712.0253}{
  [arXiv:0712.0253v3]};
\bibinfo{author}{\bibfnamefont{C.~R.} \bibnamefont{Das}} \bibnamefont{and}
  \bibinfo{author}{\bibfnamefont{L.~V.} \bibnamefont{Laperashvili}},
  \href{http://arxiv.org/abs/0712.1326}{
  [arXiv:0712.1326v1]}.

\bibitem[{\citenamefont{Jungman et~al.}(1996)\citenamefont{Jungman,
  Kamionkowski, and Griest}}]{CDMRevs}
\bibinfo{author}{\bibfnamefont{G.}~\bibnamefont{Jungman}},
  \bibinfo{author}{\bibfnamefont{M.}~\bibnamefont{Kamionkowski}},
  \bibnamefont{and} \bibinfo{author}{\bibfnamefont{K.}~\bibnamefont{Griest}},
  \bibinfo{journal}{Phys. Rept.} \textbf{\bibinfo{volume}{267}},
  \bibinfo{pages}{195} (\bibinfo{year}{1996}),
  \href{http://arxiv.org/abs/hep-ph/9506380}{
  [arXiv:hep-ph/9506380v1]};
\bibinfo{author}{\bibfnamefont{G.}~\bibnamefont{Bertone}},
  \bibinfo{author}{\bibfnamefont{D.}~\bibnamefont{Hooper}}, \bibnamefont{and}
  \bibinfo{author}{\bibfnamefont{J.}~\bibnamefont{Silk}},
  \bibinfo{journal}{Phys. Rept.} \textbf{\bibinfo{volume}{405}},
  \bibinfo{pages}{279} (\bibinfo{year}{2005}),
  \href{http://arxiv.org/abs/hep-ph/0404175}{
  [arXiv:hep-ph/0404175v2]}.

\bibitem[{\citenamefont{Hung}(2006{\natexlab{b}})}]{Hung2006b}
\bibinfo{author}{\bibfnamefont{P.~Q.} \bibnamefont{Hung}},
  \href{http://arxiv.org/abs/hep-ph/0604063}{
  [arXiv:hep-ph/0604063v4]}.

\bibitem[{\citenamefont{Hung et~al.}(2006)\citenamefont{Hung, Masso, and
  Zsembinszki}}]{Hung2006c}
\bibinfo{author}{\bibfnamefont{P.~Q.} \bibnamefont{Hung}},
  \bibinfo{author}{\bibfnamefont{E.}~\bibnamefont{Masso}}, \bibnamefont{and}
  \bibinfo{author}{\bibfnamefont{G.}~\bibnamefont{Zsembinszki}},
  \bibinfo{journal}{J. Cosmol. Astropart. Phys}
  \textbf{\bibinfo{volume}{2006}}, \bibinfo{pages}{004} (\bibinfo{year}{2006}),
  \href{http://arxiv.org/abs/astro-ph/0609777}{
  [arXiv:astro-ph/0609777v2]}.

\bibitem[{\citenamefont{Kolb and Turner}(1990)}]{Kolb1990}
\bibinfo{author}{\bibfnamefont{E.~W.} \bibnamefont{Kolb}} \bibnamefont{and}
  \bibinfo{author}{\bibfnamefont{M.~S.} \bibnamefont{Turner}},
  \emph{\bibinfo{title}{The Early Universe}}, vol.~\bibinfo{volume}{69} of
  \emph{\bibinfo{series}{Frontiers in Physics}}
  (\bibinfo{publisher}{Addison-Wesley}, \bibinfo{year}{1990}), ISBN
  \bibinfo{isbn}{0-201-11603-0}.

\bibitem[{\citenamefont{Binetruy et~al.}(1984)\citenamefont{Binetruy, Girardi,
  and Salati}}]{coann}
\bibinfo{author}{\bibfnamefont{P.}~\bibnamefont{Binetruy}},
  \bibinfo{author}{\bibfnamefont{G.}~\bibnamefont{Girardi}}, \bibnamefont{and}
  \bibinfo{author}{\bibfnamefont{P.}~\bibnamefont{Salati}},
  \bibinfo{journal}{Nucl. Phys. B} \textbf{\bibinfo{volume}{237}},
  \bibinfo{pages}{285} (\bibinfo{year}{1984});
\bibinfo{author}{\bibfnamefont{K.}~\bibnamefont{Griest}} \bibnamefont{and}
  \bibinfo{author}{\bibfnamefont{D.}~\bibnamefont{Seckel}},
  \bibinfo{journal}{Phys. Rev. D} \textbf{\bibinfo{volume}{43}},
  \bibinfo{pages}{3191} (\bibinfo{year}{1991}).

\bibitem[{\citenamefont{Scherrer and Turner}(1986)}]{Scherrer1986}
\bibinfo{author}{\bibfnamefont{R.~J.} \bibnamefont{Scherrer}} \bibnamefont{and}
  \bibinfo{author}{\bibfnamefont{M.~S.} \bibnamefont{Turner}},
  \bibinfo{journal}{Phys. Rev. D} \textbf{\bibinfo{volume}{33}},
  \bibinfo{pages}{1585} (\bibinfo{year}{1986}).

\bibitem[{\citenamefont{Gondolo and Gelmini}(1991)}]{Gondolo1991}
\bibinfo{author}{\bibfnamefont{P.}~\bibnamefont{Gondolo}} \bibnamefont{and}
  \bibinfo{author}{\bibfnamefont{G.}~\bibnamefont{Gelmini}},
  \bibinfo{journal}{Nucl. Phys. B} \textbf{\bibinfo{volume}{360}},
  \bibinfo{pages}{145} (\bibinfo{year}{1991}).

\bibitem[{\citenamefont{Dieci}(1992)}]{Dieci1992}
\bibinfo{author}{\bibfnamefont{L.}~\bibnamefont{Dieci}}, \bibinfo{journal}{SIAM
  J. Numer. Anal.} \textbf{\bibinfo{volume}{29}}, \bibinfo{pages}{781}
  (\bibinfo{year}{1992}).

\bibitem[{\citenamefont{Spergel et~al.}(2007)\citenamefont{Spergel, Bean,
  Dor{\'e}, Nolta, Bennett, Dunkley, Hinshaw, Jarosik, Komatsu, Page
  et~al.}}]{WMAP3}
\bibinfo{author}{\bibfnamefont{D.~N.} \bibnamefont{Spergel}},
  \bibinfo{author}{\bibfnamefont{R.}~\bibnamefont{Bean}},
  \bibinfo{author}{\bibfnamefont{O.}~\bibnamefont{Dor{\'e}}},
  \bibinfo{author}{\bibfnamefont{M.~R.} \bibnamefont{Nolta}},
  \bibinfo{author}{\bibfnamefont{C.~L.} \bibnamefont{Bennett}},
  \bibinfo{author}{\bibfnamefont{J.}~\bibnamefont{Dunkley}},
  \bibinfo{author}{\bibfnamefont{G.}~\bibnamefont{Hinshaw}},
  \bibinfo{author}{\bibfnamefont{N.}~\bibnamefont{Jarosik}},
  \bibinfo{author}{\bibfnamefont{E.}~\bibnamefont{Komatsu}},
  \bibinfo{author}{\bibfnamefont{L.}~\bibnamefont{Page}}, \bibnamefont{et~al.},
  \bibinfo{journal}{Astrophys. J. Suppl.} \textbf{\bibinfo{volume}{170}},
  \bibinfo{pages}{377} (\bibinfo{year}{2007}),
  \href{http://arxiv.org/abs/astro-ph/0603449v2}{
  [arXiv:astro-ph/0603449v2]}.

\end{thebibliography}
\end{document}